\documentclass{elsart4}

\usepackage{amssymb}
\usepackage[english,francais]{babel}

\newtheorem{e-proposition}[theorem]{Proposition}

\newtheorem{e-definition}[theorem]{Definition\rm}

\renewcommand{\theequation}{\arabic{equation}}

\def\og{\leavevmode\raise.3ex\hbox{$\scriptscriptstyle\langle\!\langle$~}}
\def\fg{\leavevmode\raise.3ex\hbox{~$\!\scriptscriptstyle\,\rangle\!\rangle$}}

\begin{document}


\centerline{Title of the dossier/Titre du dossier}
\begin{frontmatter}



\selectlanguage{english}

\title{Open Questions in GRB Physics}


\selectlanguage{english}

\author[authorlabel1]{Bing Zhang},
\ead{zhang@physics.unlv.edu}

\address[authorlabel1]{Department of Physics and Astronomy, 
University of Nevada Las Vegas, NV 89154, USA}

\begin{abstract}

Open questions in GRB physics are summarized as of 2011,
including classification, progenitor, central engine, 
ejecta composition, energy dissipation and particle acceleration
mechanism, radiation mechanism, long term engine activity, external 
shock afterglow physics, origin of high energy emission, 
and cosmological setting. Prospects of addressing
some of these problems with the upcoming Chinese-French GRB
mission, SVOM, are outlined.


\vskip 0.5\baselineskip




\keyword{gamma-rays: bursts; stars; black holes; neutron stars;
shocks; magnetic fields; radiation mechanism} 
}

\end{abstract}
\end{frontmatter}

\selectlanguage{english}


\section{Introduction}
\label{}

The field of gamma-ray bursts (GRBs) has rapidly advanced in 
recent years, especially following the launches of NASA missions
{\it Swift} (in 2004) and {\it Fermi} (in 2008) 
\cite{meszaros06,zhangcjaa07,gehrels09,michelson10}.
Due to their elusive nature, observing GRBs in
all wavelengths at all epochs (including during and after the
GRB) is still challenging with the current GRB detectors and
follow up telescopes. As a result, every time when a new temporal
or spectral window is unveiled, a rich trove of new phenomenology 
is uncovered. While solving some old problems, new observations 
usually raise more questions and challenges. 
This provides sustainable impetus to this still relatively
young field. In any case, current observations gradually put
together a sketch of the global picture of GRBs, although many
details remain vague or uncertain.

This review summarizes the open questions in GRB physics
as of 2011. Ten topics are discussed, including classification,
progenitor, central engine, ejecta composition, energy dissipation
and particle acceleration mechanism, radiation mechanism, long term 
engine activity, external shock afterglow physics, origin of high 
energy emission, and cosmological setting. In connection with the
upcoming Chinese-French SVOM (Space-based multi-band
astronomical Variable Object Monitor) mission, I also discuss some 
prospects of addressing some of these problems in 2015 and beyond.
More detailed discussions on the SVOM mission \cite{paul10} and
multi-wavelength observational prospects in 2015 and beyond 
\cite{greiner10} can be found in this volume.

\section{Open Questions}

\subsection{Classification}
\label{sec:classification}

In astronomy, classification is traditionally solely based on
distinct clusters in data based on well-defined criteria. Well
known examples include stellar spectral classification and supernova
classification, both having spectral line features as essential 
criteria. Since these criteria usually invoke ``yes'' or ``no''
judgments, the classification schemes of these objects are relatively
unambiguous. As transient events without any credible spectral 
feature, the GRB classification was traditionally based on their
durations in the temporal domain and hardness ratio (HR) in the spectral 
domain. An analysis of GRBs detected by BATSE (sensitive in 
30 keV - 2 MeV) suggested that there are two classes of GRBs in the
$T_{90}-{\rm HR}$ space, i.e. the long/soft
class that comprises roughly 3/4 of the population, and the short/hard
class that comprises the other 1/4 \cite{kouveliotou93}. A rough
separation line in duration, i.e. $T_{90} \sim 2$ s in the observer
frame, was suggested. 

The main issue of applying the $T_{90}$ criterion to define the
class of a GRB is that $T_{90}$ is detector dependent. GRB pulses
are typically broader at lower energies. Also a more sensitive detector
tends to detect weaker signals which would be otherwise buried in
noises. It was therefore not surprising that observations carried out 
with softer detectors such as HETE-2 and {\it Swift} brought confusions to
classification. For example, among a total 476 GRBs detected by {\it Swift} 
BAT (sensitive in 15 keV - 150 keV) from Dec. 19 2004 to
Dec. 21 2009, only 8\% have $T_{90} < 2$s \cite{sakamoto10}, much less 
than the $\sim 1/4$ fraction of the BATSE sample. An additional 2\%
of {\it Swift} GRBs have a short/hard spike typically shorter than or 
around 2s, but with an extended emission lasting 10's to $\sim 100$ 
seconds.  These bursts, dubbed
``short GRBs with E.E.'' \cite{norris06}, have $T_{90} \gg 2$s as
observed by {\it Swift}, but could be short GRBs if they were detected 
by BATSE. So the unfortunate consequence of the $T_{90}$ classification 
is that the membership to a certain category of the {\em same} GRB 
could change when the detector is changed. One possibility is to
define a burst's category based on its BATSE-band duration. Then
two issues arise: First, what is special for the BATSE band? If
other detectors such as {\it Swift} were launched earlier, what would
be the criteria to define long vs. short GRBs? Second, 
it is difficult to precisely infer $T_{90}$ in the BATSE band
using the data of other detectors (e.g. {\it Swift}/BAT). 
It requires accurate time-dependent
spectral information of the entire burst, which is only available 
for few very bright GRBs. Even for these bursts, extrapolating the
BAT band spectrum to the BATSE regime is risky and usually not
correct, since the GRB spectrum is known to be curved. 
This has been evidenced in some {\it Swift} GRBs that were co-detected 
by other detectors with harder bandpass such as Konus/Wind.  
Fortunately, the confusion in $T_{90}$ classification 
only arises in the ``grey'' area
between the two classes. For most GRBs, one can still tell
whether they are ``long'' or ``short''.

A further complication is that several groups argue that the best fit 
to the $T_{90}$ distribution histogram is three Gaussian 
functions in logarithmic space, e.g. \cite{horvath10}. This adds
one more ``intermediate-duration'' class besides the traditional
``short'' and ``long''. For the same GRB, the membership to this 
intermediate class is even more ambiguous and subject to the detector 
bandwidth and sensitivity.

The differentiation between long and short GRBs is established with
a firmer footing thanks to the afterglow and host galaxy observations.
Observations led by BeppoSAX, HETE2, and {\it Swift} suggest that at least 
some long GRBs are associated with supernova Type Ic 
\cite{galama98,hjorth03,stanek03,campana06,pian06}. Most long GRB
host galaxies are found to be dwarf star-forming galaxies 
\cite{fruchter06}. These facts establish the connection between 
long GRBs and deaths of massive stars \cite{woosley93,paczynski98}.
The breakthrough led by {\it Swift} unveiled that some nearby
short GRBs (or short GRBs with E.E.) have host galaxies that are
elliptical or early type, with little star formation 
\cite{gehrels05,fox05,barthelmy05a,berger05}. This points towards
another type of progenitor. The top candidate model for this category
is mergers of two compact objects, e.g. two neutron stars (NS-NS)
or a neutron star and a black hole (NS-BH) 
\cite{paczynski86,eichler89,narayan92}. This led to the common
ansatz that ``long GRB = massive star GRB, and
short GRB = compact star GRB''. 

Such a cozy picture was soon messed up by some observations.
GRB 060614 and GRB 060505 are two long duration nearby GRBs 
that did not have bright SN associations and that share similar
properties to short GRBs
\cite{gehrels06,galyam06,fynbo06,dellavalle06,zhang07b}. Two 
high-$z$ GRBs 090423 and 080916 have rest-frame durations shorter 
than 1s, but are likely related to massive stars
\cite{zhang09,tanvir09,salvaterra09,greiner09}. An observer-frame 
short GRB 090426 was found in many aspects similar to long GRBs
\cite{levesque10,antonelli10,xin10}.
This suggests that certain observation properties (e.g. long 
vs. short duration) do not always refer to certain types of
progenitor. 

While some appeal to modify the meaning of
``long'' and ``short'' to reflect ``massive star origin'' and
``compact star origin''\footnote{The ``long'' and ``short'' 
notations become more and more confusing, since growing data
demand to introduce more complicated notations such as 
``long short GRBs'' or ``short long GRBs''.}, 
respectively, others started to
``classify'' GRBs physically \cite{zhang07b,bloom08,zhang09},
and appeal to multiple observational criteria to determine
the ``physical category'' of a GRB \cite{donaghy06,zhang09}.
The classification here is beyond the traditional definition
of astronomical classifications (which are based on data).
Rather, they are based on some well-motivated GRB progenitor
models which are believed to be associated with GRBs (see 
more discussion below in \S\ref{sec:progenitor}). 
For cosmological GRBs that mark catastrophic explosions,
two general physical classes of GRBs (or two types of models
that are associated with GRBs) are ``massive star GRBs''
(or ``Type II GRBs'') and ``compact star GRBs'' (or ``Type
I GRBs'') \cite{zhang07b,bloom08,zhang09,kann10,kann08}. 
Since duration
alone is no longer necessarily a good indicator for the
physical category of a GRB, one must appeal to multiple
observational criteria to judge the correct physical category
of the GRB progenitor model that is associated with a certain
GRB \cite{zhang09,donaghy06}.

One may also seek for other observational parameters to conduct
GRB classification. For example, for GRBs with redshift 
measurements, a parameter $\varepsilon \equiv E_{\gamma,iso,52}
/E_{p,z,2}^{5/3}$ can be used to classify GRBs into two 
categories \cite{lv10}. The high-$\varepsilon$ and 
low-$\varepsilon$ categories are found to be more closely related 
to Type II (massive star) GRBs and Type I (compact star) GRBs,
respectively.

\subsection{Progenitor}
\label{sec:progenitor}

The progenitors of GRBs are not identified, and it will be very
difficult to identify them. There are two ways to approach this
goal. One is to use observational data to narrow down the
allowed progenitor types. The other is to use theoretical insights
to construct toy models (e.g. collapse of a massive star, merger 
of two NSs) and use analytical and numerical methods to investigate 
whether GRBs can be made. 

For massive star GRBs (typically long), the following two 
observational facts have offered important clues for the progenitor 
type: (1) A handful of GRBs are found to be associated with
supernovae Type Ic (no hydrogen lines and no or weak helium lines)
\cite{galama98,hjorth03,stanek03,campana06,pian06,starling10};
(2) The hosts are dwarf galaxies
with intense star formation, and the GRB locations
track the brightest star formation regions in the hosts 
\cite{fruchter06,savaglio09}.
Theoretically, in order to produce a relativistic jet from a 
collapsing star to power the observed GRB, one requires that
the stellar core must carry a high angular momentum
\cite{woosleyheger06}. The spin 
axis then provides a natural preferred direction for 
jet launch and propagation.

One then comes up with three requirements for a massive star GRB
progenitor: (1) These stars must track the brightest regions in
the star formation regions; (2) The hydrogen envelope is largely
depleted so that the progenitor is likely a Wolf-Rayet star; 
(3) The core carries a high angular momentum. 
Within these general constraints, several candidate progenitor
systems are possible \cite{fryer99}: collapse of a massive single
star with a high angular momentum; collapse of a massive star
in a close binary system; and merger of two He stars. For the
single star scenario, achieving both a depleted hydrogen
envelope (which requires a strong wind, and hence, high
metallicity) and a rapidly rotating core (requires low
metallicity) seem contradictory. It is argued that rapid 
mixing of H with He would result in burning H to He without
the need of ejecting the H envelope (and hence without losing angular
momentum) \cite{yoon05}. Alternatively, a binary progenitor
can retain a high angular momentum core with the H envelope
ejected \cite{woosleyheger06}.

An alternative idea to interpret massive star GRBs is to invoke
two-step explosions. Such a ``supranova'' model \cite{vietri98} 
envisages a core collapse supernova explosion weeks to months
before the GRB. This first explosion produces a rapidly rotating
massive neutron star, which subsequently collapses to form
a black hole and generates a GRB later when the centrifugal support
is not enough to hold the neutron star. The observed GRB/SN
associations suggest that the delay between the SN and GRB, if
any, cannot be more than 1-2 days 
\cite{hjorth03,stanek03,campana06,pian06,starling10}. This model
is therefore not favored for massive star GRBs.

One should keep in mind the following caveats regarding the massive
star progenitors in general. (1) Among 5-6 robust cases of GRB/SN 
associations, only GRB 030329 is a typical GRB. The rest are nearby,
low luminosity GRBs, which may form a distinct population with
a different progenitor or different central engine 
\cite{soderberg06,mazzali06,liang07}. Strictly speaking, we are
relying on one case to speculate the progenitor of most GRBs.
It is possible that all high luminosity GRBs are associated with
SNe. Observationally the SN signature is however difficult to catch, since 
these GRBs are usually not at low redshifts, and since they typically
have bright optical afterglows to outshine the SN signals.
(2) A good fraction of long GRBs, namely ``optically dark GRBs'', do not 
have a detectable optical afterglow. Their
host galaxies are usually not identified. The predominant dwarf 
star-forming host galaxies in the published sample \cite{fruchter06}
may be due to a selection effect. In fact, a {\it Chandra} 
observation of the dark GRB 090417B shows that its host galaxy
is a Milky-Way like galaxy with heavy dust extinction 
\cite{holland10}. (3) The theoretical preference of
low metallicity (to retain angular momentum) is not fully
established from the data \cite{savaglio09}, although various
arguments have been made in favor of such a condition
\cite{wolf07,li08,niino10}. 

The observational evidence for a compact star merger progenitor is 
indirect (in contrast to the massive star progenitor), which is based 
on the following observational facts: (1) GRB 050509B and GRB 050724 
are found to be associated with an elliptical host galaxy (the former)
or an early type host galaxy with low star formation rate (the latter)
\cite{gehrels05,bloom06,barthelmy05a,berger05}; (2) Statistically, 
most identified short GRB hosts are late type galaxies, but
GRBs track less bright regions of the galaxies, and have larger
offsets from the centers \cite{fong10};
(3) Deep searches placed stringent upper limits on supernova light
associated with several nearby short GRBs
\cite{bloom06,fox05,berger05,hjorth05,kann08}.
Theoretically, compact star mergers would have much less and much
denser fuel than massive star core collapses, which would power short
duration GRBs \cite{rosswog03,aloy05,rezzolla11}.

The two popular models involving compact star mergers include NS-NS
mergers \cite{paczynski86,eichler89,narayan92} and NS-BH mergers
\cite{paczynski91}. The double NS systems are known to exist in our
galaxy, and their orbits are known to shrink due to gravitational
radiation \cite{taylor89,kramer08}. They are doomed to merge 
someday, so the NS-NS merger model has been the earliest 
cosmological model for GRBs. The NS-BH binary systems are not
observed. This does not necessarily mean that they are much rarer 
than NS-NS systems (although they should be rarer). This is because
at least one pulsar is expected to be ``recycled'' in the NS-NS 
system, so that its radio beam is much wider than normal PSRs 
and detections are eased. The pulsar in the BH-NS system, however,
is not expected to go through this recycling phase, so that the
pulsar detectability is significantly lower. In any case,
both merger models are argued to be able to interpret many short 
GRB properties (see \cite{nakar07,lee07} for reviews).

Despite the indirect supports to the merger scenarios mentioned
above, following observational facts raise cautions to take the
merger progenitors as truth. First, a group of short GRBs are found 
to have redshifts $>1$ \cite{berger07}, some having luminosity
typical for long GRBs \cite{deugartepostigo06}. Some apparent long
GRBs, including the two highest-$z$ GRBs 
\cite{tanvir09,salvaterra09,greiner09}, have rest-frame duration 
shorter than 1 s. The high energetics of these GRBs 
demands extremely narrow beaming of the NS-NS progenitor, which 
is difficult to achieve theoretically. Numerical modeling consistently
reveals wide jets in merger systems due to the lack of a dense medium
(e.g. the stellar envelope as expected in massie star core collapse
scenarios) to help to collimate the jet \cite{aloy05,rezzolla11}.
Only BH-NS systems powered by BH spin may 
power these energetic events,
but these mergers (with a massive BH and rapid BH spin) are very
rare. Second, in order to account for many high-$L$ short GRBs and
not-too-many low-$L$ nearby short GRBs, the luminosity function
of short GRBs should be shallow \cite{virgili10}. Since most 
merger models predict a redshift-distribution that peaks at
low-$z$, the shallow luminosity function is translated 
to a shallow peak flux distribution ($\log N - \log P$), which
violates the BATSE short GRB sample constraint significantly \cite{virgili10}.
The consistency between the merger models and the {\it Swift} $z$-known
sample \cite{nakar06} or the BATSE sample \cite{guetta06} was
claimed shortly after the discovery of short GRB afterglows.
However, the {\it Swift} sample was too small and the two samples
({\it Swift} vs. BATSE) were not jointly considered in those analyses.
A recent joint analysis shows a sharp inconsistency between various
merger models and the observational constraints ($z-L$ distributions
and $\log N - \log P$ distributions) \cite{virgili10}. This may
suggest that either the merger models are not the correct model 
for short GRBs, or not all short GRBs are from compact star mergers.
A recent modified compact star coalescence scenario is to invoke dynamical 
``collisions'' of NS-NS or NS-BH systems in globular clusters
\cite{lee10,rosswog11}. If this population of coalescences dominates
over mergers in the field galaxy due to gravitational wave
radiation, the above conflict may be alleviated.
In any case, since NS-NS and NS-BH merger models 
predict specific gravitational
wave signals \cite{cutler02}, a definite test to the merger models 
of short GRBs may be achieved in the future when gravitational wave
detections become possible. 

Besides mergers, several other types of progenitor have been proposed
for short GRBs. One scenario is accretion-induced-collapses
(AICs) of NSs (e.g. \cite{dermer06}), which is similar to the 
``supranova'' model for long GRBs \cite{vietri98}, but with a much
longer delay between the SN and GRB. Cosmological short GRBs may
be produced this way.
Soft Gamma-ray Repeaters (SGRs) produce giant flares with a short,
hard spike, which would be recognized as short/hard GRBs in nearby
galaxies \cite{palmer05,hurley05,tanvir05}. It is likely that the
short/hard GRB population is contaminated by these events, but the
fraction of contamination is believed to be small \cite{nakar06b}.

\subsection{Central Engine}

Different types of progenitor may result in a common central engine
that powers the observed GRBs. 
Observations suggest that a GRB central engine should satisfy
the following requirements: (1) It can drive an outflow with 
extremely high luminosity and energy. If the emission is
isotropically distributed in all directions, the required jet
luminosity ranges from $L_{\rm iso} \sim 10^{47}-10^{54}~
{\rm erg~s^{-1}}$, and the total gamma-ray energy ranges from
$E_{\rm iso} \sim 10^{49} - 10^{55}~{\rm erg}$ 
\cite{fishman95,zhangmeszaros04}; (2) The ejecta
need to be ``clean'' with small baryon contamination, so that they can
achieve a relativistic speed, with Lorentz factor $\Gamma$ typically
greater than 100 \cite{piran99,lithwick01,liang10}, some even close
to 1000 \cite{abdo09a,abdo09b,abdo09c}; (3) The outflow needs 
to be collimated, with a beaming factor $f=\Delta \Omega/4\pi \sim 1/500$ 
for bright GRBs \cite{frail01,bloom03,liang08,racusin09}, so that
the real luminosity and energy of a GRB is reduced by this factor;
(4) The engine needs in general to be intermittent, with a range of 
variability time scales \cite{fishman95,beloborodov98}. In some GRBs, 
the engine can generate smooth  (but varying) lightcurves 
\cite{norris02,campana06}; (5) The engine can last long,
with renewed, progressively less powerful late activities to power 
X-ray flares and other activities 
(see \S\ref{lateengine} for full discussion).
 
Several types of GRB central engine have been discussed in the 
literature. The leading candidate is a black hole (possibly rapidly 
spinning) + torus system. An alternative candidate is a rapidly 
spinning, highly magnetized NS (magnetar). A more exotic possibility 
is a compact star solely composed of quark matter, i.e. a quark star.
There are three energy reservoirs involved in these engines:
the accretion power, the spindown power of the central object,
and the phase transition power. 

The black hole - torus engine is widely discussed in both the 
collapsar scenario \cite{woosley93,macfadyen99,proga03,zhangw03},
and the compact star merger scenario \cite{narayan92}.
The first energy source is the accretion power from the torus.
Neutrino annihilation from the torus can drive a hot jet along
the spin axis to power the GRB.
The accretion powered jet luminosity is $L_{\rm acc} = \zeta 
\dot M c^2 \sim 1.8\times 10^{51}~{\rm erg~s^{-1}} \zeta_{-3}
(\dot M/1 M_\odot~ {\rm s^{-1}})$. In order to achieve the observed 
GRB luminosity, the accretion rate should be close to (0.1-1) 
$M_\odot /{\rm s}$ 
for a reasonable efficiency factor $\zeta$ to convert accretion power 
into jet power. The second energy source is the spin energy of
the black hole. This energy can be tapped by magnetic
fields threading the ergosphere of a Kerr black hole through
the Blandford-Znajek (BZ) mechanism
\cite{blandford77,meszarosrees97b,lee00,li00,mckinney05}. The jet luminosity 
$L_{\rm BZ} \simeq 0.1 B^2 (\Omega_{\rm BH}^2/c)(GM_{\rm BH}/c^2)^2 
\simeq 3\times 10^{50}~{\rm erg~s^{-1}} B_{15}^2 (M/3 M_\odot)^2
a^2 f(a)$, where $a$ is the dimensionless spin parameter of the black
hole, and $f(a)$ is a increasing function of $a$. In order to power
a GRB, the black hole should be rapidly spinning ($a \lesssim 1$), the
accretion rate should be high so that magnetic fields near the 
horizon are strong enough (the radial magnetic field strength near 
the black hole is $B \gtrsim 10^{15}$ G). For a same $a$, 
a more massive BH (and hence, more spin energy to tap)
would give a more luminous burst. Such a BZ-powered jet
would carry a strong magnetic field, and is likely Poynting-flux-dominated.
It is possible that both mechanisms (neutrino annihilation
and BZ process) are operating in BH-torus systems. 
A variable outflow can be due to the interplay between the magnetic
fields and the accreting materials \cite{proga03b,lei09}. For a
jet emerging from a star, jet propagation instabilities in the 
envelope can give rise to further variabilities in the outflow
\cite{zhangw03,morsony10}.

The main power of a millisecond magnetar engine is its spindown
power \cite{usov92}, which is $L_{\rm sd} = B^2 \Omega^4 R^6/6c^3
\sim 3.7\times 10^{50}~{\rm erg~s^{-1}} B_{15}^2 \Omega_4^4 R_6^6$
for the dipole spindown model.
In order to power a GRB, the magnetar must have a surface magnetic 
field $B \gtrsim 10^{15}$ G, and an angular frequency $\Omega \gtrsim 10^4$
Hz (which corresponds to a spin period $P \lesssim 0.6$ ms). Notice that
$\Omega \sim 10^4$ is already close to the upper limit of the angular
frequency of a neutron star. 
The maximum total energy of a magnetar engine is defined
by its spin energy $E_{spin} \sim (1/2)I \Omega^2 \sim 5\times 10^{52}$
erg. Increasing $B$ would reduce the spindown time scale $\tau_{sd} = 
3c^3 I/(B^2 R^6 \Omega^2)\simeq 800 ~{\rm s} I_{45} B_{15}^{-2} R_6^{-6} 
\Omega_4^{-2}$. In principle, the engine cannot power
an intrinsically luminous and long GRB. One may invoke a small beaming
factor $f$ to accommodate a large isotropic luminosity. However, for
millisecond rotators the open field lines have a large solid angle, 
so that $f$ cannot be too small unless another medium (e.g.
envelope or supernova ejecta) serve to collimate the jet.
It is possible that the accretion 
power also operates in a magnetar. The neutrino annihilation rate is 
enhanced with respect to the BH-torus system \cite{zhangdai09}. 
However, given the high accretion rate (e.g. $1 M_\odot/{\rm s}$), 
the NS would quickly (e.g. in 2 seconds) turn into a BH. Another issue 
is that a neutrino-driven wind from a proto neutron star tends to be 
``dirty'' \cite{janka95,qian96}, which cannot produce a clean fireball
to power a GRB. A GRB may be generated after the proto neutron star
cools and a Poynting flux dominated outflow is launched, typically
several seconds later \cite{thompson04,metzger07,metzger11}. Scenarios
to have a magnetic bubble penetrating through the stellar envelope 
without significant contamination have been discussed 
\cite{wheeler00,lyutikov03}. 

Strange quark matter could be more stable than neutron matter 
\cite{witten84}, so that strange quark stars could form in high
pressure environments for a wide range of allowed parameters for 
QCD \cite{alcock86}. A quark star engine has been invoked to power 
a GRB in various contents 
\cite{chengdai96,dailu98a,ouyed05,paczynski05,xuliang09}.
There are two advantages of introducing a quark star engine.
First, extra energy sources due to phase transitions (from
neutron matter to 2-flavor quark matter, from 2-flavor to 3-flavor
strange quark matter, and quark matter condensation 
\cite{chengdai96,ouyed05,xuliang09}) are introduced. Second, since 
the star is bound by strong interaction rather than gravity,
neutrinos and photons can be released without launching materials
to contaminate the fireball \cite{chengdai96,ouyed05,paczynski05}.
The time scale for phase transitions
may be fast. In order to launch a highly variable jet, intermittent
accretion is needed, and the engine power includes both the
accretion power and the phase transition power \cite{ouyed05}.

Since all three types of engine are argued to satisfy most 
observational constraints, identifying the right one among them 
using observational data is not straightforward. Among the three 
possible engines, the BH - torus system is most naturally expected. 
For massive star GRBs, studies of Type Ic SNe associated with some 
GRBs suggest a large enough mass for the progenitor star to form a
BH rather than a NS \cite{iwamoto98,mazzali03}. A BH-torus 
engine is relevant for BH-NS mergers. For
NS-NS mergers, the total mass of the two NSs ($\sim 2.8 M_\odot$)
is believed to exceed the maximum NS mass for most NS equation of
state, so that a BH - torus engine is also likely. Nonetheless,
evidence of a NS (QS) engine in some GRBs is collected. First,
spectral modeling of SN 2006aj associated with GRB 060218 suggests 
that this SN has much smaller ejecta mass and kinetic energy
than other Type Ic SNe associated with GRBs, pointing towards
a massive star progenitor that is not massive enough to produce a 
BH \cite{mazzali06}. Second, the spin down luminosity of a NS (QS) 
should have a constant luminosity plateau followed by a 
$L\propto t^{-2}$ decay. This signature may show up in the
early afterglow phase. The continuous injection of pulsar
spindown energy onto the blastwave would result in a 
shallow decay phase in the early afterglow phase 
\cite{dailu98b,zhangmeszaros01a}, which may account for the
plateau feature in the early afterglows of some {\it Swift} GRBs
\cite{zhang06,nousek06,obrien06}. If the pulsar wind has strong
dissipation before landing on the blastwave and if the engine
ceases suddenly (probably due to collapse into a black hole), 
an ``internal plateau'' would appear in the X-ray afterglow, 
characterized by a plateau phase followed by a very steep decay 
as observed in some {\it Swift} GRBs \cite{troja07,liang07b,lyons10}
(see \S2.7 for more discussion). We therefore
suggest that although BH - torus systems may be common in most
bright, energetic GRBs, pulsar systems may exist in at least
some GRBs. Unfortunately, besides theoretical arguments,
there is essentially no ``smoking-gun'' observational criterion
to differentiate a QS engine from a NS engine. 

A dedicated review on numerical simulations of GRB central
engines can be found in this volume \cite{novak10}.

\subsection{Ejecta Composition}

The composition of a GRB outflow includes three components:
matter, magnetic fields, and photons. Photons are advected 
with matter and magnetic fields initially, and are decoupled 
from the ejecta at the photosphere radius, where Compton
scattering optical depth drops below unity. Above the 
photosphere, the jet carries a matter flux and a magnetic flux,
which is essentially a Poynting flux in the lab frame 
because of the existence of an induced electric field. 
More photons are generated from the regions where kinetic
energy or magnetic energy is dissipated (i.e shocks or magnetic
reconnection regions), which escape the
ejecta without further coupling. The distribution of energy
between matter and magnetic fields is denoted by the
magnetization factor $\sigma \equiv B^2/4\pi\Gamma \rho c^2$
(where the magnetic field $B$ and the matter density 
$\rho$ are measured in the lab frame), the
ratio between Poynting flux and matter flux. Within the
matter content, one has the baryonic and leptonic
components. The relative distribution can be denoted by
a parameter $Y=n_\pm/ n_p$, the number ratio between leptons and 
protons. Usually the baryonic component dominates in mass unless  
$Y \geq m_p/m_e$, the proton-to-electron mass ratio. 
Due to the extreme temperature (typically $k T \sim $ MeV) and 
density (typically nuclear density) at the central engine, 
heavy ions are less likely to survive in the jet, so the 
dominant charged baryons
in the jet are protons. For both compact star merger and massive star 
core collapse central engines, it is likely that a noticeable fraction
of free neutrons exist in the fireball, and initially are coupled
with protons through strong interaction 
\cite{derishev99,beloborodov03}. These neutrons decouple from
the ion ejecta, decay with a comoving life time $\sim 900$ s,
and would leave interesting observational features 
\cite{derishev99,beloborodov03b,fan05b}. The abundance of free 
neutrons is usually denoted by the neutron-to-proton number ratio, 
$\xi \equiv n_n/n_p$.

The traditional GRB models are built in the matter-dominated regime
($\sigma \ll 1$). This is the standard ``fireball'' shock model
\cite{paczynski86,goodman86,shemi90,rees92,meszarosrees93,rees94}.
Magnetic fields are likely to be entrained in the ejecta. In
the matter-dominated models, they are believed not to play a 
kinematically dominant role. As $\sigma$ approaches and exceeds 
unity, magnetic fields become kinematically important. In such
a magnetically-dominated jet, the ejecta 
would carry a globally ordered magnetic field. Notice that giving
a same total outflow luminosity and at a same distance from the 
central engine, the absolute strength of magnetic fields does not 
vary significantly from $\sigma \lesssim 1$ to $\sigma \gg 1$. 
This is because the Poynting flux does not differ significantly, 
and different $\sigma$ is mostly caused by different mass flux 
of the flow. Another comment is that due to magnetic acceleration 
and dissipation, $\sigma$ is expected to drop with radius
\cite{drenkhahn02,vlahakis03,komissarov09,tchekhovskoy09,zhangyan10}.
As a result, one needs to specify a radius (from the central engine)
when judging whether the flow is matter-dominated or 
magnetically-dominated. 

Diagnosing the composition of GRB ejecta has not been an easy task. 
Although the $\sigma \ll 1$ models have been widely discussed
mostly because of their simplicity, evidence that magnetic fields
are playing an important role at least in some GRBs is gradually
accumulating. (1) If $\sigma \geq 1$, an ordered magnetic field
would give rise to strong linearly polarized synchrotron emission
\cite{waxman03,lyutikov03b,granot03,toma09b}.
Strong linear polarization of gamma-ray emission has been claimed
in some GRBs (e.g. \cite{coburn03,willis05}), although the results
are subject to large uncertainty \cite{rutledge04}. 
(2) Recent {\it Fermi} observations of GRB 080916C \cite{abdo09a} revealed
a series of nearly featureless Band-function spectra covering 6-7
orders of magnitude in energy throughout the entire burst. Such
an observational fact brings challenge to the traditional fireball
internal shock model. If the observed Band component is the non-thermal
emission from internal shocks, one would expect a bright
quasi-thermal spectral component from the fireball photosphere
which outshines the non-thermal component, making the observed
spectrum significantly deviated from the simple Band form. This
led to the suggestion that the ejecta has to be Poynting flux
dominated in order to suppress the bright photosphere thermal 
emission \cite{zhangpeer09,fan10}. Since most {\it Fermi} LAT GRBs have
Band-only spectra similar to GRB 080916C \cite{zhang10}, one may
speculate that most GRBs may function similar to GRB 080916C. 
The $\sigma \ll 1$ dissipative photosphere model 
(e.g. \cite{beloborodov10,lazzati10}) could give
rise to a Band-like spectrum, but the model predictions cannot
reproduce the broad spectra of GRB 080916C (e.g. \cite{zhangyan10,zhang10}
for detailed discussion). 
The {\it Fermi} LAT GRB 090902B \cite{ackermann10} is a special case that 
shows bright quasi-thermal emission in the time-resolved spectra
\cite{ryde10,zhang10}, which can be well interpreted as the photosphere
emission \cite{peer10}. The magnetization parameter $\sigma$ is likely
not very high, but the magnetic fields need to be strong in any 
case \cite{peer10}. (3) As the ejecta is decelerated by the ambient
medium, the existence and strength of the reverse shock depends on
the GRB composition \cite{fan04b,zhangkobayashi05,mimica09,mizuno09}.
The general trend is the following \cite{zhangkobayashi05}: 
when $\sigma \ll 1$, the reverse shock (RS) emission becomes 
progressively stronger as $\sigma$ increases, because the synchrotron
emission becomes progressively stronger in a stronger magnetic field. 
The RS brightness
reaches the peak around $\sigma \sim 0.1$. When $\sigma$ gets close
and surpasses unity, the strong pressure from the magnetic field 
compensates part of the forward shock (FS) thermal pressure, so that the
RS becomes progressively weaker until eventually disappears when the
magnetic pressure can fully balance the FS pressure. Studying the strength
of the RS can therefore diagnose ejecta composition. The
bright optical flashes seen in several GRBs (e.g. GRBs 990123, 021211,
061126) require that the RS region is much more magnetized than the FS
region, suggesting that the engine is carrying a strong magnetic field
\cite{fan02,zhang03,kumar03,gomboc08}, with a $\sigma$ close to (but does
not exceed) unity. An early optical polarimetry observation of GRB 090102
revealed a $10\pm 1 \%$ polarization degree of emission during the
early steep decay phase believed to be of the RS origin \cite{steele09}.
This suggests that the central engine ejecta carried an ordered magnetic
field. 

Besides the above observational diagnostics, claims about the GRB
composition may be made using indirect theoretical modeling. For example,
different models predict different radii of gamma-ray emission.
The photosphere radius is typically $R_{ph} \sim (10^{11}-10^{12})
$ cm from the central engine. Internal shocks, on the other hand, occur
at distances $R_{\rm IS} \sim \Gamma^2 c \delta t \sim 3\times 10^{13}
~{\rm cm} \Gamma_{2.5}^2 \delta t_{-2}$, where $\delta t$ is
the typical variability time scale. Magnetic dissipation may occur
in various radii. For models that invoke a striped-wind field geometry
(relevant to pulsar-like central engines), significant magnetic dissipation 
can occur below the photosphere, so that the photosphere emission is
enhanced. For models invoking helical magnetic geometry (relevant to 
black hole - torus engines), significant magnetic dissipation may not 
easy to occur at small radii, but rather occur at a large
enough radius where the ordered field lines are distorted enough
so that field lines with opposite orientations can approach each other
and reconnect. Fast magnetic dissipation may be 
triggered either by collision-induced turbulent reconnection
\cite{zhangyan10}, by a switch from the collisional to
the collisionless reconnection regimes \cite{mckinney11}, or by
current instability \cite{lyutikov03}. In all these
cases, the dissipation radius is usually larger than the photosphere
radius, typically with $R_{mag} \geq 10^{14}$ cm. Finally, for a 
neutron rich outflow, neutrons decay in all radii, but
with a characteristic decay radius $R_{n} \sim 900 c \Gamma = 2.7 \times
10^{15}~{\rm cm} \Gamma_2$. Measuring the location of the MeV GRB 
emission $R_{\rm GRB}$ may shed light into the unknown composition
of the GRB ejecta.

Observationally, it is not straightforward to 
measure $R_{\rm GRB}$ using the MeV data. Nonetheless,
there are three indirect ways to infer this radius using X-ray,
optical and GeV emission data. (1) {\it Swift} GRBs typically show a rapidly 
decaying early X-ray afterglow, which is found to be connected to prompt
gamma-ray emission \cite{tagliaferri05,barthelmy05b,zhangbb07}. 
The leading interpretation is that this is the high-latitude emission
of a conical jet after the prompt emission ceases abruptly
\cite{kumar00,zhang06,liang06,zhangbb09}. Within this interpretation, 
the duration of the steep decay phase is defined by $\Delta t_{steep}
\sim (R_{\rm GRB}/c) (1-\cos\theta_j) (1+z) \sim (R_{\rm GRB}/c) 
(\theta_j^2/2) (1+z)$. For a typical jet angle $\theta_j 
\sim 0.1$, the data generally require that $R_{\rm GRB} \geq 10^{15}$
cm \cite{lyutikov06,kumar06}; (2) Some GRBs have prompt optical
emission detected roughly tracking gamma-ray emission 
\cite{vestrand05,racusin08}. If this optical emission is from the
same emission region as gamma-rays, the GRB emission site can be 
constrained by requiring that the synchrotron self-absorption
frequency is below the optical band. For the naked-eye GRB 080319B,
this gives $R_{\rm GRB} \sim 10^{16}$ cm \cite{racusin08,kumarpanaitescu08}.
Constraints from other GRBs with prompt optical detections or upper
limits give $R_{\rm GRB} \geq ~{\rm several} \times 10^{14}$ 
cm for typical GRB Lorentz factors \cite{shen09}; (3) {\it Fermi} observations 
suggest that for a good fraction of GRBs, the MeV Band-function spectra 
extend to the GeV range, suggesting that GeV emission is from the same 
region as the MeV emission. For these bursts, the detected maximum GeV 
photon energy can be used to constrain $R_{\rm GRB}$ along with
$\Gamma$ \cite{gupta08}. For example, GRB 080916C gives a constraint
$R_{\rm GRB} \geq 10^{15}$ cm in general \cite{zhangpeer09}, and 
$R_{\rm GRB} \sim 10^{16}$ cm if $R_{\rm GRB} \sim \Gamma^2 c \delta t$ 
is assumed \cite{abdo09a}. A general picture emerging from these
indirect constraints is that $R_{\rm GRB}$ is usually large, typically
$\sim 10^{15}$ cm. This is also consistent with some model
constraints based on MeV observations \cite{kumar07,kumarmcmahon08}.
These large emission radii 
are consistent with the expectation of high-$\sigma$ models, although 
low-$\sigma$ models are not ruled out (but the parameter space is
constrained). The caveat here is that the optical/GeV emission may
not always be from the same region as MeV emission. For example, various
arguments suggest that the gamma-ray emission radius may be smaller
than that of optical emission in GRB 080319B \cite{fan09,zou09,resmi10}.
The distinct GeV component of GRBs 090902B and 090510 is very likely
from a different radius from the MeV component 
\cite{abdo09b,ackermann10,zhang10}.

To summarize, the case of GRB composition is inconclusive. Evidence of 
a strongly magnetized central engine is accumulating, although the 
$\sigma$ value in the GRB emission region is not well constrained.
It is possible that $\sigma$ may
vary from burst to burst. This may sound unnatural. However, as
explained above, when $\sigma$ is greater than, say, 0.3, the
increase in $\sigma$ does not correspond to a further increase
of Poynting flux, but rather corresponds to a decrease in the
associated matter flux. As a result, a slight change in matter
flux may result in a significant change in the $\sigma$ value
in the outflow (e.g. from $\sigma \lesssim 1$ to $\sigma \gg 1$).
This is entirely possible.
Since $\sigma$ is a decreasing function of radius
\cite{drenkhahn02,vlahakis03,komissarov09,tchekhovskoy09,zhangyan10},
one needs to specify a radius of reference for comparison in order 
to get a coherent picture of GRB magnetization. For example,
it is possible that $\sigma \gg 1$ initially in the GRB emission
region, $\sigma$ is moderately high in the GRB emission region,
and $\sigma \lesssim 1$ at the deceleration radius
after the global magnetic dissipation process is over \cite{zhangyan10}.

GRB composition is greatly tied to two interesting topics: whether
GRBs are the dominant sources of ultra-high energy cosmic rays (UHECRs) 
\cite{waxman95,vietri95} and high energy neutrinos
\cite{waxman97,meszaroswaxman01,razzaque04b}.
These models all invoked a baryon-dominated outflow. If GRBs are 
on average Poynting-flux-dominated, the strengths of these
signals would drop by a factor $(1+\sigma)^{-1}$, rendering GRBs not 
necessarily the dominant contributors to UHECRs and the high energy 
neutrino background.

\subsection{Energy Dissipation and Particle Acceleration Mechanism}

The main energy sources of a GRB include the gravitational accretion 
energy ($E_{\rm grav} \sim (1/2)G M m/R \sim 1.3\times 10^{53} ~{\rm erg} 
(M/10 M_\odot) (m/1 M_\odot) / R_7$, where $M$ is the mass of the accretor 
and $m$ is the mass of accreted materials) and the rotation energy 
of the central BH ($E_{\rm rot,BH} \sim M_{\rm BH} c^2
[1-\sqrt{(1/2)(1+\sqrt{1-a^2})}] \leq 0.29 M_{\rm BH} c^2
\sim 5.2 \times 10^{54}~{\rm erg} (M_{\rm BH}/10M_\odot)$) or
NS ($E_{\rm rot,NS} \sim (1/2) I \Omega^2 \sim 5\times 10^{52}~{\rm erg} 
I_{45}\Omega_4^2$). Besides,
the central engine carries a magnetic field energy
$E_{\rm mag} \sim (1/6) R^3 B^2 \sim 1.7\times 10^{50} B_{15}^2 R_7^3$.
For quark star scenarios, an extra phase transition energy (of the same
order of the accretion energy) may be added to the energy budget.

In a GRB, energy is transferred among various forms in different stages.
A fraction of the gravitational potential energy is initially converted 
into thermal energy, forming a hot fireball of photons, electron/positron
pairs and a small number of baryons. The fireball expands under its
own thermal pressure, and converts thermal energy to the bulk kinetic
energy of the ejecta. Torqued by magnetic fields, the 
spin energy of the central object can be converted into a Poynting flux, 
which is entrained in the ejecta. The ejecta can be also accelerated
under the internal magnetic pressure of the ejecta. 

At the photosphere radius, photons initially advected in the fireball 
are released, giving rise to a quasi-thermal spectrum (probably
modified by Compton upscattering). This is the first location 
where photons are released.
Above the photosphere, kinetic energy can be converted into particle 
energy and then into radiation in shocks. Alternatively, Poynting 
flux energy can be converted into particle energy and then into
radiation in reconnection regions. These dissipation regions are
additional sites to emit photons.  The observed GRB emission
is from one or more of these emission sites.

Shock dissipation is the widely discussed energy dissipation 
mechanism. The internal collisions within an unsteady matter-dominated
wind injected from the central engine give rise to internal shocks
\cite{rees94,paczynski94}.
The relative Lorentz factor between the two colliding shells can
range from mildly relativistic ($\Gamma_{rel} \gtrsim 1$) to relativistic
($\Gamma_{rel} \sim $ a few to tens). After the collisions, it is
assumed that shells merge. The leading fast shell interacts with
the circumburst medium and drives an ``external'' forward shock
into the medium \cite{rees92,meszaros93}. A reverse shock propagates 
into the ejecta until crossing it 
\cite{meszarosrees97,saripiran99,kobayashi00}. 
The shocked materials between the forward and the reverse shocks
form a ``blastwave''.
The forward shock is initially relativistic. The reverse shock is
mildly relativistic if the central engine duration is not long
(the thin shell regime), but could be highly relativistic if the 
central engine duration is long enough (the thick shell regime) 
\cite{sari95}. During the self-similar deceleration phase
\cite{blandford76}, the blastwave may be refreshed by slow
ejecta lagging behind \cite{rees98} or a Poynting flux injected
by a long-lasting central engine (e.g. a spinning down millisecond
pulsar or magnetar) \cite{dailu98b,zhangmeszaros01a}.

Particles (both baryons and leptons) are believed to be accelerated 
in shocks. The well known process is the first-order Fermi acceleration.
The effect is well known in the non-relativistic regime. For
relativistic shocks, particle-in-cell (PIC) simulations are 
starting to unveil the acceleration details 
\cite{nishikawa05,spitkovsky08,nishikawa09}. A power
law tail develops as simulation time grows. A relativistic 
Maxwellian component is still observed in the current simulation, 
although it may be significantly eroded eventually \cite{spitkovsky08}.
Observationally there is no significant signature for such a
relativistic Maxwellian component. 

For a magnetized upstream, the shock jump condition is modified
\cite{kennel84}. The energy that is available for dissipation
is the matter part, which is $1/(1+\sigma)$ of the total energy
\cite{zhangkobayashi05}. Without magnetic dissipation, the
magnetic energy in the flow (a portion of $\sigma/(1+\sigma)$) 
remains intact. For deceleration of a magnetized ejecta
by an unmagnetized circumburst medium, the strength of the
reverse shock progressively decreases \cite{zhangkobayashi05,mimica09}
until completely disappears as $\sigma$ increases to 10s
to $\sim 100$, when the forward shock internal pressure is
no longer stronger than the magnetic pressure in the ejecta
\cite{zhangkobayashi05}. For collisions between two magnetized
shells, a pair of shocks would propagate into both shells as
long as the ram pressure exceeds the magnetic pressure of 
the shells. These shocks are weak, but in any case, would
serve to distort the ordered magnetic field lines in the shell
\cite{zhangkobayashi05,zhangyan10}.

Whether and how particles are accelerated in magnetized shocks is 
subject to more investigations. PIC simulations \cite{sironi09a}
suggest that for a relativistic ($\Gamma \gtrsim 5$) magnetized 
($\sigma > 0.03$) shock, particle acceleration is possible only 
within a narrow range of magnetic inclination angles ($\lesssim
34^{\rm o}/\Gamma$). On the other hand, the reverse shock model
to interpret early optical flashes require that the reverse shock 
is more magnetized than the forward shock (with $\sigma$ close to 
0.1) \cite{zhang03,fan02,kumar03}. A $\sim 10\%$ linear
polarization degree was measured in the early optical afterglow
for GRB 090102 \cite{steele09}, which is consistent with emission
from a magnetized reverse shock. The inconsistency between
data constraints and PIC simulations may be partially alleviated
by the following two factors: First, for thin shells the reverse 
shock invoked in GRB modeling is usually transrelativistic (i.e.
reaching mildly relativistic at the end of shock crossing)
\cite{sari95}. So the allowed magnetic inclination angle space
is much larger. Second, If the outflow was initially Poynting
flux dominated, it must have gone through significant magnetic
dissipation during the GRB prompt emission phase so that $\sigma$
has dropped to around or below unity. Turbulence
may have significantly distorted the ordered magnetic field
configuration, so that during the deceleration phase, the
magnetic field lines are no longer mostly perpendicular to the
shock normal \cite{zhangyan10}.

Besides shocks, magnetic dissipation is another effective way
to accelerate relativistic particles. This is associated with
reconnection of magnetic field lines with opposite polarity.
The full details of reconnection is not well understood. The
main difficulty has been that reconnection speed is slow in
a steady state \cite{sweet58,parker57}. In the GRB context,
continuous slow reconnection below and slightly above the 
photosphere may enhance photosphere emission, and lead to an
up-scattered non-thermal tail above the photosphere thermal
peak \cite{thompson94,rees05,thompson07,lazzati10}. In order
to produce bursty emission above photosphere to power the
observed GRBs, the traditional Sweet-Parker reconnection speed
is too low. A possibility is that reconnection proceeds
via turbulence, so that multiple sites reconnect simultaneously 
\cite{lazarian99}. For a high-$\sigma$ flow, turbulence may be
induced through multiple internal collisions which distort the
ordered magnetic fields. A reconnection-turbulence cascade may
result, which would discharge a significant fraction of magnetic
energy to power a GRB with high radiative efficiency. This is
the Internal Collision-induced MAgnetic Reconnection and 
Turbulence (ICMART) model of GRBs \cite{zhangyan10}.
Particle acceleration in such turbulence-reconnection events is 
a difficult problem. Qualitatively, one can have three competing 
processes: direct electric field acceleration in current sheets, 
first-order Fermi acceleration between two approaching magnetic
field lines, and the second-order Fermi acceleration in
turbulence. Unfortunately, no currently available numerical tools 
can model the detailed particle acceleration processes in a 
relativistic, turbulent, dissipative, and strongly magnetized 
fluid. An alternative proposal for fast reconnection is the
switch between collisional and collisionless regimes for a striped
wind geometry \cite{mckinney11}. It is interesting to further 
investigate how this proposal may account for the GRB prompt 
emission phenomenology.

Another particle acceleration mechanism in a strongly magnetized
flow is the comoving Poynting flux acceleration \cite{liangedison09},
or inductive and electrostatic acceleration \cite{smolsky96,ng06}.
Particles ``surf'' on a wave of electric fields and gain energy.
The applications of such a mechanism to GRBs have been discussed
but are not explored in detail.

A dedicated review on particle acceleration in GRBs 
can be found in this volume \cite{lemoine10}.

\subsection{Radiation Mechanism}

GRB prompt emission is characterized by a smoothly-joint
broken-power-law named ``Band'' function \cite{band93}. The 
``non-thermal'' nature of the spectrum demands that the emission
is produced by a population of particles (likely leptons) with
a power law energy distribution. The leading radiation mechanisms
include synchrotron (or jitter) radiation, synchrotron self-Compton 
(SSC), and Compton upscattering of a thermal seed photon source.
GRB afterglow has a broad-band spectrum at any epoch. The main
contributor to afterglow emission is the
synchrotron and SSC emission of the electrons accelerated
in the external shock. 

While the origin of the external shock afterglow is relatively well 
modeled (see \S2.8 for further discussion), the radiation mechanism
of the prompt GRB emission is subject to debate.

The leading mechanism is synchrotron radiation 
\cite{meszaros94,tavani96}. This is because it is the most 
naturally expected non-thermal emission mechanism. 
The GRB central engine is likely magnetized. Internal shocks 
can also generate magnetic fields through plasma instabilities 
\cite{medvedev99,nishikawa09}.
Shock accelerated electrons must gyrate in the magnetic fields and
radiate synchrotron photons. 

The most straightforward synchrotron model, however, suffers a
list of criticisms. (1) The synchrotron cooling time scale is
typically much shorter than the dynamical time scale, so that
electrons are in the ``fast cooling'' regime. The expected photon
spectrum below $E_p$ is supposed to have a photon index $\alpha
=-1.5$, while the observations show a typical value of $\alpha
\sim -1$. This is the fast cooling problem \cite{ghisellini00}.
The possible solutions to this problem include introducing
a rapidly decaying magnetic field in the internal shock region
\cite{peerzhang06}, introducing slow heating in the emission
region (e.g. in an ICMART event) \cite{zhangyan10,asano09}, 
and introducing Klein-Nishina cooling  
\cite{daigne10}. (2) The hardest low energy
photon index is supposed to be $\alpha = -2/3$ in the synchrotron
model (corresponding to the $F_{\nu} \propto \nu^{1/3}$ regime
of synchrotron emission. However, a fraction of GRBs have 
$\alpha$ even harder than this ``synchrotron line of death''
\cite{preece98}. Possible solutions include introducing 
contributions from the thermal photosphere \cite{meszarosrees00},
considering synchrotron self-absorption \cite{lloyd00}, and
introducing ``jitter'' radiation \cite{medvedev00}.
(3) For typical parameters, the predicted $E_p$ from synchrotron
emission is about 2 orders of magnitude smaller than the observed
value. This requires that only a small fraction of electrons
are accelerated in the internal shocks \cite{daigne98,bykov96}.
The same requirement is needed to correctly derive the 
synchrotron self-absorption frequency in internal shocks
\cite{shen09}. In a high-$\sigma$ flow, this problem is naturally 
solved \cite{zhangyan10}. (4) Within the internal shock model,
it is argued that the allowed parameter regime for synchrotron
model is greatly constrained \cite{kumarmcmahon08}. Such a
limitation applies to the internal shock model. For magnetic
dissipation synchrotron models such as the ICMART model 
\cite{zhangyan10}, the constraint is much weaker. 

A variant of the synchrotron radiation mechanism is the
jitter radiation mechanism \cite{medvedev00}. Within this 
scenario, magnetic fields have too small a coherence length 
$\lambda_B$ so that electrons cannot make a complete gyration. 
The typical jitter emission frequency no longer depends on the
strength of the magnetic field, but is related to $\lambda_B$. 
The low energy photon index below $E_p$ can range from 0 to -1
\cite{medvedev06}. One issue of this scenario is that the 
assumed small coherence scale is not revealed from PIC numerical
simulations of relativistic shocks \cite{sironi09b}. On the
other hand, such a small coherence scale may be realized in
magnetic reconnection regions \cite{zenitani08}, 
so that the radiation mechanism
may be relevant in models that invoke magnetic dissipation as
the origin of GRB emission.

The second radiation mechanism candidate to interpret prompt
GRBs is SSC. Within this scenario, the synchrotron radiation
peaks in the IR/optical/UV range, and the observed GRBs are
dominated by the SSC emission \cite{panaitescu00b}. It was found
that within the internal shock model, the allowed parameter space
of SSC is much larger than that of synchrotron if all the electrons
are accelerated \cite{kumarmcmahon08}. Introducing an assumption
that only a small fraction of electrons are accelerated would
largely alleviate this problem \cite{bosnjak09}.
The SSC mechanism attracted serious attention following the 
discovery of the ``naked-eye''
GRB 080319B \cite{racusin08}. Bright optical pulses are found
associated with gamma-ray pulses, with flux greatly exceeding
the extrapolation of gamma-ray emission into the optical band.
An immediate possibility is that optical is due to synchrotron
radiation, while gamma-rays are due to SSC 
\cite{kumarpanaitescu08,racusin08,kumarnarayan09}. Counterarguments 
against the simplest SSC mechanism for GRB 080319B 
include the energy crisis (since $Y\sim 10$,
the 2nd order IC component would be even more energetic) 
\cite{derishev01,zou09,piran09},
$\sim 1$ s lag of the optical emission with respect to the gamma-ray
emission \cite{beskin10}, and the difficulty of interpreting the more 
variable gamma-ray lightcurve (than optical) \cite{resmi10}.
More complicated SSC models invoking turbulence in the emission 
region may overcome some of these criticisms \cite{kumarnarayan09}.

The third radiation mechanism commonly discussed in the literature
is Compton upscattering of thermal photons. The leading scenario
is upscattering off thermal photons from the jet photosphere
\cite{thompson94,rees05,peer06,thompson06,thompson07,giannios08,beloborodov10,lazzati10,peerryde10}. This requires energy dissipation below 
and slightly above the
photosphere. Within this scenario, the observed $E_p$ is essentially
the temperature of the photosphere. Upscattering naturally gives
rise to a power law spectrum above $E_p$. The spectral index
below $E_p$, is however difficult to alter from the Rayleigh-Jeans
slope (corresponding to $\alpha=+1$ rather than $\alpha=-1$ as
typically observed). Considering sub-photosphere heating and
equal-arrival-time effect in a relativistic ejecta, one gets 
$\alpha \sim 0.4$ \cite{beloborodov10}, 
still much harder than the observed
spectrum. The equal arrival time effect would modify $\alpha$
to approach $-1$ at late times when the flux already drops
significantly \cite{peerryde10}. This may be relevant when the central
engine activity is over, but is not applicable during the prompt
emission phase when a continuous wind is ejected from the central
engine. In general, the dissipative photosphere models have the
difficulty to interpret the low energy photon spectral index of GRBs.

Besides the photosphere, other thermal sources can give rise to
seed photons for relativistic electrons to up-scatter. These
photon sources include the thermal photons released from the
exploding star \cite{lazzati00}, the thermal photon ``glory''
from the progenitor star that is trapped by the environment
\cite{dar04}, as well as thermal photons from shock breakout
\cite{wang07}.

Finally, hadronic mechanisms (proton synchrotron and proton-photon
interaction to produce charged and/or neutral pions) have been
also suggested to interpret prompt GRB emission 
(e.g. \cite{gupta07,asano09b,razzaque10}.
Since protons are radiatively inefficient as compared with
leptons, these models require that protons carry most of the
total energy (i.e. proton dominated).

It is difficult to identify the correct radiation mechanism with
the current available data. The ``smoking gun'' would be the
gamma-ray polarization data. Although both synchrotron radiation 
in an ordered magnetic fields \cite{lyutikov03,granot03} and IC
viewed at certain angles \cite{shaviv95,lazzati04} can give high 
degree of polarization, systematically studying the statistical 
polarization properties of a sample of GRBs 
can differentiate the competing
models and may lead to identification of the radiation mechanism
of GRB prompt emission \cite{toma09b}.

A dedicated review on multiwavelength GRB prompt emission 
observations can be found in this volume \cite{atteia10}.

\subsection{Long Term Central Engine Activity}\label{lateengine}

One of the major discoveries of {\it Swift} is that the so-called
``afterglow'' is not only the emission from the external shock 
(as generally believed in the pre-{\it Swift} era), but also includes 
emission from a long-lasting central engine. The arguments for
a long lasting central engine lie in the following three
pieces of evidence.

The most convincing evidence is the existence of X-ray flares
following nearly half of {\it Swift} GRBs 
\cite{burrows05,chincarini07,falcone07}. These flares have
sharp rise and sharp decay. The morphology is essentially
impossible to be interpreted within the framework of the 
external shock model \cite{ioka05}. One therefore needs to
appeal to late central engine activities to interpret them
\cite{burrows05,zhang06,fanwei05,lazzati07,maxham09}. The
most convincing evidence of such an interpretation is the
reset of the clock for each episode of engine activity, as 
is suggested from the data: in order to interpret the decay
following an X-ray flare as the high latitude emission effect, 
the required $T_0$ is usually right before the rise of the 
flare \cite{liang06}. Such a property cannot be naturally 
interpreted within other X-ray flare models that do not 
invoke late central engine activities (e.g. \cite{beloborodov10b}). 

Next, a small fraction of GRB X-ray afterglows show an extended
plateau followed by a sudden drop of flux with a decay index
steeper than 3 (sometimes even as steep as 9) 
\cite{troja07,liang07b,lyons10}. Such a rapid fall cannot 
be interpreted within the external shock models, and needs
to invoke internal dissipation of a long lasting central engine
wind. Such a plateau is therefore sometimes also called an
``internal plateau'' \cite{liang07b,lyons10}. For comparison,
most X-ray afterglow lightcurves show
a plateau followed by a ``normal'' decay with slope $\sim -1$
\cite{zhang06,nousek06}. Within the external shock interpretation,
the X-ray plateaus are due to adding energy into the blastwave.
Before observing the internal plateaus, there have been two
possibilities to account for such a refreshed shock: a long
lasting central engine \cite{dailu98b,zhangmeszaros01a} or
piling up slow materials ejected promptly \cite{rees98,sarimeszaros00}.
With the discovery of internal plateaus, it is now clear that
some GRBs indeed have a long-lasting central engine activity,
so that the former scenario can operate at least in some GRBs.

Last, some authors even interpret the most common X-ray plateaus
(those followed by a $t^{-1}$ normal decay phase) as emission
from the central engine \cite{ghisellini07,kumar08b,cannizzo09,lindner10}.
The main reason is that in more than half cases, no optical
break was discovered at the transition time from the plateau
to the normal decay phase \cite{panai06b,liang07b}, and 
that there is essentially no spectral evolution across the
temporal break \cite{liang06b}. Such a chromatic behavior is
very difficult to interpret within the framework of the external
shock models. Invoking two-component external shock jets
\cite{depasquale09} would require contrived shock parameters
and unnatural assumptions. Invoking a long lasting reverse shock
\cite{genet07,uhm07} would require that the forward shock emission
is suppressed by many orders of magnitude. Another possibility,  i.e.
dust scattering effect \cite{shao05}), can interpret the lightcurve
but not the spectral evolution as observed for most GRBs \cite{shen09b}. 
One therefore is left with the possibility that the X-ray emission is
dominated by the internal emission of a long-lasting central engine.
Indeed, a quantitative X-ray afterglow model attributing X-ray emission
to the dissipative photosphere of a long lasting central engine wind
can reproduce the shallow, normal, and late break of X-ray lightcurve
with chromatic features as observed \cite{wu10}.

To conclude, the observational data demand the following two aspects
related to the GRB central engine. First, the engine is usually 
still active after the prompt emission phase is over, and can last 
up to $> 10^4$ s after the trigger. Second, there can be two types
of central engine activities, an erratic component that can power 
X-ray flares, and a smooth component that can power the plateaus
(those followed by a steep decay segment or even those followed
by a normal decay segment). 

There is no well accepted mechanism yet to account for the 
erratic behavior of the late central engine activities. Nonetheless,
some ideas have been proposed. These include fragmentation of the
collapsing star \cite{king05}, fragmentation of the accretion disk
of both massive star and compact star GRBs \cite{perna06}, intermittent
accretion behavior caused by a time variable magnetic barrier 
\cite{proga06}, magnetic bubbles launched in a post-merger 
differentially rotating, proto-neutron star \cite{dai06}, 
helium synthesis in the post-merger debris of a compact star
GRB \cite{lee09}, and quakes in solid quark stars \cite{xuliang09}. 
In any case, since the neutrino-driven jet
luminosity drops steeply when the accretion rate becomes small
\cite{popham99}, the existence of low-luminosity X-ray flares
at late times demand that the late jets that power X-ray flares
are driven and collimated by magnetic mechanisms \cite{fan05e}.
Another caveat is that an intermittent jet may not always correspond
to intermittent accretion. It may be possible that accretion is
continuous, but another mechanism (e.g. magnetic
activity) is responsible for the observed intermittent jet emission
(e.g. \cite{yuan10}).

The mechanisms to account for a long-lasting continuous late central 
engine activity include fall-back accretion onto the black hole engine
\cite{macfadyen01,rosswog07,lazzati08}, or tapping the spin energy
of a strongly magnetized millisecond pulsar \cite{dailu98b,zhangmeszaros01a}.
For the latter possibility, one needs to interpret the energetic prompt
gamma-ray emission without introducing a black hole - torus central
engine. Possible mechanisms include hyperaccretion onto a neutron
star \cite{zhangdai09} or phase transition into a quark star 
\cite{chengdai96,dailu98a,ouyed05,paczynski05}.

To firmly test the long-lasting central engine scenarios, additional
observational channels are needed. For example, X-ray polarization
measurements for X-ray flares would be essential to test the magnetic
nature of the late jet \cite{fan05e}. Gravitational wave observations
during the plateau phase would shed light onto the nature of the
millisecond pulsar central engine \cite{corsi09}.

\subsection{External Shock Afterglow Physics}

Among all the topics discussed in this review, the external shock
afterglow physics is the most definite one. Even though the forward 
shock afterglow model has some free parameters\footnote{For the 
standard model, five parameters are needed: the total energy of
the blastwave $E_{\rm iso}$, one parameter to describe the circumburst 
medium - the number density $n$ for a constant ISM model, or the 
$A_\star$ parameter for the stellar wind model, one parameter for
electron energy distribution $p$, and two parameters for shock
energy partition - the electron equipartition parameter $\epsilon_{\rm e}$
and the magnetic field equipartition parameter $\epsilon_{\rm B}$.}, the
dynamics of the blastwave, the radiation spectrum at any instant,
as well as the lightcurve at any wavelength can be uniquely 
predicted once these input parameters are specified 
\cite{meszarosrees97,sari98,dailu98c,chevalier00}. As a result,
using multi-band (radio, optical and X-rays) data, one can constrain
these unknown parameters based on the model 
\cite{wijers99,panaitescu01,panaitescu02,yost03}. More complicated
factors, such as jet break \cite{rhoads99,sari99}, energy injection
into the blastwave \cite{rees98,dailu98b,zhangmeszaros01a,zhangmeszaros02a},
angular structure of the jet 
\cite{meszaros98,zhangmeszaros02b,rossi02,kumar03,racusin08}, as 
well as transition to the non-relativistic phase \cite{huang99,huang03}
can be added into the model to interpret more complicated 
observational behaviors.

Modeling afterglow using the external shock model
has been the main topic before {\it Swift}. {\it Swift} observations of
early afterglows on the other hand revealed a more complicated
picture. The X-ray afterglows show a canonical lightcurve with
5 distinct components \cite{zhang06,nousek06}. The early steep
decay component connecting the prompt emission 
\cite{tagliaferri05,barthelmy05b,zhangbb07} (Component I of
\cite{zhang06}) and the erratic X-ray flares (Component V of
\cite{zhang06}) are believed to be of the internal origin.
The rest three segments: shallow decay (II), normal decay (III)
and the late steep decay (IV) are well interpreted within the
external shock model: II is the external shock during the 
continuous energy injection phase, III is the external shock
emission after injection is over, and IV is the external shock
emission after the jet break phase \cite{zhang06,nousek06,panai06a}. 
According to such an interpretation, the two breaks are hydrodynamical
breaks and should also appear in other wavelengths as well (i.e.
achromatic). This was indeed seen in some GRBs (e.g. GRB 060729 and
GRB 060614 \cite{grupe07,mangano07b}). However, chromatic behaviors
were discovered around both breaks in a good fraction of GRBs
\cite{panai06b,liang07b,liang08}. This suggests that the X-ray
afterglow might not be always of the 
external shock origin. Therefore, in the
{\it Swift} era, the afterglow modeling made one step backwards: instead
of constraining the afterglow model parameters using the data, one
needs first to identify which emission component is from the 
external forward shock, and which is not.

An important result from the pre-{\it Swift} era was that massive star 
(Type II) GRBs are collimated, and the jet opening angles are such
distributed that the jet corrected energy is roughly constant 
\cite{frail01,bloom03,berger03}. {\it Swift} observations raise cautions
to accept this conclusion readily. There is no platinum jet breaks
identified (i.e. those clearly show achromatic feature in all
wavelengths) \cite{liang08}. Some bursts have chromatic breaks
at the so-called ``jet break'' epoch, while some bursts do not have a
break at much late epochs \cite{liang08,racusin09}. More late time
optical observations (e.g. \cite{dai08}) are needed to reveal 
jet breaks and GRB collimation and energetics.

Another important aspect of the external shock afterglow physics is
the emission of the reverse shock that propagates into the ejecta. 
Traditionally the ejecta is approximated as a finite width shell 
with uniform density. The reverse shock therefore gives a short-live
emission signature. Since the density of the shell is approximately
$\Gamma$ times of that of the medium, and since the FS and RS regions
share roughly the same pressure and internal energy, each electron
in the RS region has less energy than those in the FS, so that the
peak synchrotron frequency is smaller by roughly $\Gamma^2$. As
a result, while the FS emission initially peaks in X-rays and above,
RS emission peaks in IR/optical/UV 
\cite{meszarosrees97,saripiran99,kobayashi00}. 
At the shock crossing time, the synchrotron emission peak fluxes,
typical frequencies between RS and FS are connected through some
simple relations \cite{kobayashizhang03a,zhang03}. For typical
shock parameters for both the FS and the RS, the RS emission is not 
expected to be very bright. The optical lightcurve is characterized
by double peaks, the first RS peak marking fireball deceleration,
and the second FS peak related to crossing of the typical synchrotron
frequency in the optical band \cite{zhang03}.
The bright optical flashes characterized by a $F_\nu \propto t^{-2}$
decay followed by a normal $F_\nu \propto t^{-1}$ decay 
(e.g. \cite{akerlof99}) requires that the RS is more magnetized
than the FS \cite{zhang03,gomboc08}. There is also a regime where
the typical RS synchrotron frequency is way below the optical band,
so that there is essentially no RS signature in the optical 
lightcurve \cite{jinfan07}. Considering ejecta magnetization,
the RS flux increases initially as $\sigma$ increases from below,
but would start to decrease as $\sigma$ approaches unity and 
gradually diminishes when $\sigma \gg 1$ \cite{zhangkobayashi05,mimica09}.
A group of GRBs are found to have a smooth afterglow onset bump
dominated by the emission from the FS (e.g. \cite{liang10}).
These either have a low RS typical frequency \cite{jinfan07} or have
a highly magnetized ejecta at the deceleration time 
\cite{zhangkobayashi05,mimica09}.

For a wind medium, the RS is usually relativistic, which is
accompanied by a prominent optical emission signature
\cite{wu03,kobayashizhang03b}. 
SSC in the RS region \cite{kobayashi07} or cross IC between the
electrons and photons from FS and RS \cite{wang01,wang01b} could
be important when the RS emission is prominent, 
which may contribute to early X-ray and gamma-ray
afterglow emission.

It is possible that the ejecta do not have a uniform Lorentz factor,
luminosity and density. Considering an ejecta with a more complicated
stratification profile, the RS emission lightcurve can be made to 
have rich features, including reproducing the canonical X-ray
lightcurve as observed by {\it Swift} \cite{genet07,uhm07}. However,
for nominal parameters, the FS emission is brighter by several
orders of magnitude. In order to interpret the data with the RS
model, one has to argue that the FS emission is suppressed
\cite{genet07,uhm07}.

A dedicated review on GRB afterglow
can be also found in this volume \cite{godet10}.

\subsection{Origin of High Energy Emission}

Before {\it Fermi}, emission above 100 MeV was detected only in a handful
of GRBs (e.g. \cite{hurley94,gonzalez03}). The Large Area Telescope 
(LAT) on board {\it Fermi} makes it possible to detect high energy emission 
from GRBs regularly. Although somewhat lower than the pre-launch 
predictions, the current detection rate of $\sim 9$ per year
allows collection of a moderate sample of GRBs with detected
emission above 100 MeV
\cite{abdo09a,abdo09b,abdo09c,ackermann10,zhang10}.
A dedicated review on {\it Fermi} LAT GRB science
can be also found in this volume \cite{piron10}.

The properties of GRB high energy emission can be summarized as
follows (e.g. \cite{zhang10} for a comprehensive study): (1) The
LAT band emission is found delayed with respect to the GBM band 
emission in several (but not all) GRBs 
\cite{abdo09a,abdo09b,ackermann10,zhang10}. (2) Among
17 LAT GRBs detected before May 2010, 14 have time-resolved spectra
consistent with being a single Band function extending to both low
energy and high energy regimes \cite{zhang10}. An example is 
GRB 080916C \cite{abdo09a}, which shows a featureless Band function
covering 6-7 orders of magnitude. The Band function parameters
(the low- and high-energy photon indices $\alpha$ and $\beta$ and
peak energy $E_p$) do not vary significantly as the time resolution
becomes progressively smaller. This suggests that the Band function
may not be a superposition of many more elemental spectral components
(e.g. blackbody), but is an elemental spectral component of its 
own \cite{zhang10}. (3) Three GRBs have at least two spectral
components in the time resolved spectra. GRB 090510 and GRB 090902B
have an extra power law component extending to high energies
\cite{abdo09b,ackermann10,zhang10}. GRB 090926A may have an extra
spectral component setting in after $\sim$ 11 s, which may have a 
high energy spectral cutoff \cite{ackermann11}. (4) The MeV component of 
GRB 090902B is likely of the thermal origin. The time integrated 
spectrum is a narrow Band function plus a power law. As the time
bin becomes progressively smaller, the time-resolved spectra of
the MeV component become
progressively narrower \cite{zhang10}, so that it can be fit with
a multi-color blackbody function \cite{ryde10} or a blackbody 
function \cite{zhang10} (with the superposition with a power law
component). The MeV component of GRB 090510 can be fit as a 
power law with an exponential cutoff \cite{abdo09b,zhang10},
which may have a similar origin as the MeV component of GRB 090902B.
(5) Overall, there might be three elemental spectral components
in the prompt GRB spectra \cite{zhang10}: a Band component (Band), 
a thermal-like component (BB), and an extra power law component 
with a possible cutoff at high energies above the LAT band (PL). 
The observed GRB spectra may be decomposed of one or more of these 
elemental components. Current observed combinations include
Band only (e.g. GRB 080916C), BB + PL (e.g. GRB 090902B), and 
probably Band + PL (e.g. GRB 090926A). Other possible combinations
include Band + BB and Band + BB + PL, which exist in some GRBs
and may be tested with future bursts. 
(6) During the prompt emission phase (defined by
GBM-band emission), high energy photons are usually found to
track the MeV photons in time. This not only is valid for the 
Band-only GRBs such as GRB 080916C, but also applies to GRBs
showing a distinct spectral component at high energy (e.g. GRB
090902B) \cite{zhang10}. This hints that during the prompt emission
phase, the high energy photons are likely of an internal origin.
(7) Puzzlingly, the $>100$ MeV photons decay more slowly than
the MeV photons. After the GBM-band burst is over, LAT-band
photons are usually observed to decay with a single power law
with a slope $\sim -1.4$ \cite{ghisellini10,zhang10}. This 
suggests either that the entire high energy emission has a 
different origin from the MeV emission (which is in contrast
with the points 2 and 6 above) \cite{kumar09,kumar10,ghisellini10},
or that the high energy lightcurve is the superposition
of two components with a second (external shock) component 
setting in as the prompt emission fades \cite{zhang10,maxham11}. 

Even though the origin of high energy emission is still subject
to debate, these new data shed light into several open questions
in GRB prompt emission physics (e.g. topics discussed in \S2.4,
\S2.5 and \S2.6). (1) The featureless Band-only GRBs
such as GRB 080916C are difficult to interpret within the 
simplest baryonic fireball picture. If the entire non-thermal
spectrum is from the internal shocks, then the photosphere
emission of the hot fireball is expected to be bright enough
to show up above the detected Band spectrum \cite{zhangpeer09,fan10}.
This led to the suggestion that GRB 080916C and most LAT GRBs
have $\sigma \gg 1$ at the central engine, and probably also
in the emission region as well \cite{zhangpeer09}. 
The bright thermal emission of GRB 090902B, on the other hand,
point towards a fireball picture \cite{peer10}, with $\sigma
\lesssim 1$. This would suggest a diverse composition among
GRBs. The possibility that
the Band spectrum is from a dissipative photosphere has been 
discussed (e.g. \cite{toma10,beloborodov10,lazzati10}). These
models have specific predictions that are not consistent with
the data. For example, the predicted spectrum cannot 
extend to energies higher than $\sim 1$ GeV, while the Band 
spectra extend all the way to rest-frame $\sim 70$ GeV 
for GRB 080916C. The assumption that this prompt GeV emission
is from the external shock is not supported by detailed data
analysis \cite{zhang10}. Also the low energy photon index of a 
dissipative photosphere is predicted to be $\alpha \sim 0.4$,
which is much harder than the observed $\alpha \sim -1$. 
(2) The long-term GeV emission may be originated from
the external shock. This requires some extreme parameters for 
the external shock \cite{kumar09,kumar10}, a radiative 
blastwave \cite{ghisellini10} or a Klein-Nishina cooling
dominated shock \cite{wang10,feng10}. GeV emission during the prompt
emission phase, however, is not easy to interpret within
the external shock model \cite{he10,liu10,maxham11}, and is likely 
of an internal origin, as suggested by the data \cite{zhang10}.
(3) The delayed onset of GeV emission has been interpreted as
emergence of the upscattered cocoon emission \cite{toma09},
synchrotron emission from shock accelerated protons 
\cite{razzaque10}, delayed residual internal shock 
emission \cite{li10}, as well as the delayed fireball
acceleration to an extremely high Lorentz factor
\cite{ioka10}. Alternatively, it can be simply
due to change of particle acceleration condition or pair
production opacity during the early stage of a GRB \cite{zhangyan10,zhang10}.

Bright GRBs co-detected by {\it Fermi}/LAT and {\it Swift} would be highly
valuable to understand the nature of high energy emission.
This is because a {\it Swift} BAT trigger would lead to early XRT
and UVOT observations. Some GeV models (e.g. the external shock
model) have specific predictions in the X-ray and UV/optical 
band. The detections of early afterglows by {\it Swift} would prove
or disprove these predictions, so that one can narrow down the 
allowed models for GRB high energy emission. Unfortunately, 
the chance of LAT/{\it Swift} joint trigger is low. The only case so
far (GRB 090510) was a short GRB \cite{depasquale10}. More
cases, especially for long GRBs, are highly desirable.

\subsection{Cosmological Setting}

GRBs are cosmological events. The redshift distribution of GRBs
spans from $z=0.0085$ (for GRB 980425 \cite{galama98})
to $z=8.2$ (for GRB 090423 \cite{tanvir09,salvaterra09}). 
Several open questions of GRBs within the cosmological context 
include the following:

Are massive star GRBs good tracers of star formation
history of the universe? Do these GRBs favor a low metallicity
environment? Does the GRB luminosity function evolve with time? 
These questions are related to each other. In order to account
for the detections of GRB 080916 at $z=6.7$ \cite{greiner09} and 
GRB 090423 at $z=8.2$ \cite{tanvir09,salvaterra09}, the GRB event
rate at high-$z$ should be higher than the simple extrapolation
of the known star formation history (e.g. \cite{hopkins06}) to 
higher redshifts \cite{kistler08,salvaterra09}. 
One possibility is that GRBs still follow the
star formation history of the universe, but there is a rise of 
SFR at high $z$ due to the contribution from the population III 
stars \cite{bromm06}. Although the high-$z$ GRB excess may be 
interpreted this way, the fact that the observed high-$z$ GRBs are
not different from their nearby sisters \cite{tanvir09,salvaterra09}
suggest that this factor is not adequate to interpret the current
data. One therefore needs to argue that GRBs favor low metallicity
environment \cite{li08,butler10,qin10,virgili10b} or their luminosity
function evolves with redshift so that luminosity is higher at 
higher-$z$. These two effects are coupled with each other, and may
not be differentiated with the $L-z$ and $\log N - \log P$ data
\cite{virgili10b,qin10}. On the other hand, the low metallicity
possibility is favored by the host galaxy modeling as well
(e.g. \cite{wolf07,campisi09,niino10}).

Can high-$z$ GRBs probe the reionization history of the universe? 
The universe is known to be re-ionized around $z\sim 6$. The 
details of the reionization history is not well constrained
(e.g. \cite{holder03}). GRBs are believed to exist at as early 
as $z\sim 20$. As bright beacons in the ``dark ages'', these
high-$z$ GRBs can probe the cosmic reionization history in 
their near IR afterglow spectrum. In particular, the damping wing
bluewards of the Lyman-$\alpha$ ``Gunn-Peterson'' trough carries
the information of the neutral hydrogen column density along the
line of sight. This can in principle extend the intergalactic
medium (IGM) ionization state mapping (previously by quasars 
\cite{fanx06}) to higher $z$'s. One issue is that the GRB host
damped Lyman-$\alpha$ (DLA) system would contribute to the observed
neutral column, so that the IGM absorption feature is not clean
(e.g. \cite{totani06}). Numerical simulations suggest that the 
GRB host DLA column decreases with redshift \cite{nagamine08,pontzen10}.
Such a feature makes it more promising to use GRBs to probe
reionization at $z > 8$. Since afterglow is rapidly fading, 
very early IR observations for high-$z$ GRBs are needed to make
breakthrough in this direction.

Do high-$z$ GRBs have a different progenitor from the low-$z$ ones?
The first generation stars may be more massive than the nearby
massive stars \cite{abel02}. If these massive stars produce GRBs,
they should be powered by supermassive black holes through the
Blandford-Znajek mechanism \cite{komissarov10,meszarosrees10}. 
These GRBs should be more powerful than normal GRBs. No such 
energetic high-$z$ GRBs have been detected so far. On the other
hand, recent numerical simulations revealed that the halo to form
first generation stars may fragment, so that the first generation
stars may be in binary systems with smaller masses \cite{turk09}.
The GRBs in such systems may be then not much different from
the normal GRBs. 

Observationally, the two highest-$z$ GRBs (080916 and 090423) 
both have a rest-frame duration shorter than 2 s. One possibility
is that this is due to a selection effect (i.e. the fainter pulses
are buried below the noises). However, if the effect is intrinsic,
one would wonder whether they are massive star (Type II) GRBs
or compact star (Type I) GRBs (e.g. \cite{zhang09} and references
therein). Applying the multiple criteria besides the duration
and hardness information, it is highly likely that these bursts
are Type II GRBs \cite{zhang09}. Under certain conditions (e.g.
slow rotators), it is possible that a massive star only has a
small torus after prompt collapse, so that a short duration GRB
can be powered \cite{janiuk08a,janiuk08b}. If future high-$z$
GRBs prefer short durations, one then needs to seriously address
why the conditions for a short accretion time are preferred for
high-$z$ massive stars.

A dedicated discussion on GRBs as cosmological probes
can be also found in this volume \cite{petitjean10}.

\section{The SVOM connection}

The Chinese-French mission SVOM  is a multi-wavelength
GRB observatory scheduled to launch in 2014-2015 \cite{paul10}.
It carries four space-flown instruments: a wide field X-ray/soft
$\gamma$-ray (4-250 keV) detector ECLAIRs, a hard gamma-ray
(50 keV - 5 MeV) detector GRM, a visible telescope VT, and
an X-ray telescope MXT. A set of three ground based dedicated
instruments, including two robotic telescopes (GFTs) and one
wide angle optical monitor (GWAC), will complement the space
borne instruments. Its operation window overlaps with that
of {\it Fermi} and {\it Swift} and probably overlaps with other planned
GRB missions (such as JANUS, EXIST) as well. After 2014-2015,
several multi-messenger detectors (e.g. neutrino detector
Icecube and gravitational wave detector Advanced LIGO) will
be fully operating. It is foreseen that an exciting era of
GRB study will be ushered in. 

It is unrealistic to solve all the open questions discussed
in this review, but some aspects of the problems will be
better addressed for sure in the SVOM era. Here I discuss
some prospects.

\begin{itemize}
\item {\bf Classification \& Progenitor:} 
ECLAIRs and GRM can give independent
$T_{90}$ measurements for many GRBs. This will give a large sample
to study energy-dependent $T_{90}$ classification. Similar to
{\it Swift}, SVOM will lead broad-band follow-up observations for many
GRBs. This will allow collection of multi-criteria needed
to diagnose the physical origin of the GRBs (e.g. \cite{zhang09}). 
SVOM will continue to study massive star (Type II) GRBs and 
compact star (Type I) GRBs to allow better understanding of their
progenitors. SVOM is not ideal to detect standard short GRBs,
but the sensitivity of ECLAIRs to soft X-rays 
would help to detect more cases of
``extended emission'' of short GRBs. This would offer a chance
to study the condition of extended emission, as well as whether
short GRBs with extended emission are different from the canonical
short GRBs. SVOM will also be
powerful to study nearby low-luminosity long GRBs, allowing
a better statistics for these events, addressing their supernova
associations, event rate, luminosity function, as well as whether
they indeed form a distinct population from canonical GRBs.
\item {\bf GRB Prompt Emission Physics: }
ECLAIRs and GRM cover a different spectral window from {\it Fermi}
GBM and LAT. The joint spectral analysis between the two
instruments would allow diagnose of the prompt GRB emission
spectral components (e.g. \cite{zhang10}). In particular,
it allows a systematic analysis of the X-ray excess in the spectrum,
addressing the existence/strength of the photosphere thermal
emission component, shedding light into jet composition,
energy dissipation, particle acceleration and radiation
mechanisms. GWAC will also regularly monitor prompt optical
emission. Detections or upper limits of optical emission during
the prompt phase would lead to the constraint on the broad-band
prompt spectrum, and therefore nail down the radiation mechanism 
(e.g. synchrotron vs. SSC) of the prompt GRB emission.
\item {\bf Afterglow and ``foreglow'' physics:}
VT will regularly record early optical afterglow emission from
GRBs. Being redder and deeper than {\it Swift} UVOT, VT is ideal to
address the early optical emission physics, including categorizing
the early afterglow behavior and addressing the origin of optically
dark GRBs. Together with MXT, VT can address 
chromatic/achromatic behaviors (for temporal breaks and flares)
between the two energy bands. This would allow a systematic
study of the long-term central engine activity of GRBs.
GWAC will be monitoring the SVOM field of view during and
even before the GRB trigger. This provides a chance of studying
prompt optical emission and leading to detection of upper limits
of optical emission before the prompt emission. In view of the
recent motivation of discussing GRB ``prior emission''
(e.g. \cite{yamazaki09,liang09}), both positive
and negative detections of these ``foreglows'' would shed light
onto the function of the GRB central engine.
\item {\bf Cosmological setting:}
Having a softer bandpass than {\it Swift}/BAT, 
ECLAIRs may be able to detect more 
soft high-$z$ GRBs. Deep upper limits of VT below 0.95$\mu$m would
promptly provide good GRB candidates with $z>6$. So ideally SVOM
would lead to more detections and identifications of high-$z$
GRBs, offering the opportunity to better address high-$z$
star formation history, metallicity effect, GRB evolution, 
cosmic reionization, as well as whether population III stars
could give rise to GRBs.
\end{itemize}


\section*{Acknowledgements}
I acknowledge extensive discussion on the SVOM science with 
Jian-Yan Wei, Yu-Lei Qiu, Jin-Song Deng, and other members of 
the Chinese SVOM team, as well as with Frederic Daigne, En-Wei Liang, 
and Bin-Bin Zhang. I also thank Peter M\'esz\'aros and 
two referees for helpful comments.
This work is supported by NSF through grant
AST-0908362, and by NASA through grants NNX10AD48G, NNX09AT66G,
NNX10AP53G, and NNX08AE57A.


\bibliographystyle{ee}
\bibliography{ms}

\begin{thebibliography}{100}
\expandafter\ifx\csname url\endcsname\relax
  \def\url#1{{\tt #1}}\fi
\expandafter\ifx\csname urlprefix\endcsname\relax\def\urlprefix{URL }\fi

\bibitem{meszaros06}
P.~{M\'esz\'aros}.
\newblock {Gamma-ray bursts.}
\newblock Reports of Progress in Physics 69 (2006) 2259--2322.

\bibitem{zhangcjaa07}
B.~{Zhang}.
\newblock {Gamma-Ray Bursts in the Swift Era}.
\newblock Chinese Journal of Astronomy and Astrophysics 7 (2007) 1--50.

\bibitem{gehrels09}
N.~{Gehrels}, E.~{Ramirez-Ruiz}, D.~B. {Fox}.
\newblock {Gamma-Ray Bursts in the Swift Era}.
\newblock \araa 47 (2009) 567--617.

\bibitem{michelson10}
P.~F. {Michelson}, W.~B. {Atwood}, S.~{Ritz}.
\newblock {Fermi Gamma-ray Space Telescope: high-energy results from the first
  year}.
\newblock Reports on Progress in Physics 73 (2010) 7 074901.

\bibitem{paul10}
J.~{Paul}, J.~{Wei}, S.~{Basa}, {S. N.}~{Zhang}.
\newblock {The SVOM mission}.
\newblock this volume.

\bibitem{greiner10}
J.~{Greiner}, A.~{Rau}.
\newblock {The multiwavelength context in 2015 and beyond}.
\newblock this volume.


\newblock {Fermi Gamma-ray Space Telescope: high-energy results from the first
  year}.
\newblock Reports on Progress in Physics 73 (2010) 7 074901.


\bibitem{kouveliotou93}
C.~{Kouveliotou}, et~al.
\newblock {Identification of two classes of gamma-ray bursts}.
\newblock \apjl 413 (1993) L101--L104.

\bibitem{sakamoto10}
T.~{Sakamoto}, et~al.
\newblock {The second Swift BAT gamma-ray burst catalog}.
\newblock \apj (2011) submitted


\bibitem{norris06}
J.~P. {Norris}, J.~T. {Bonnell}.
\newblock {Short Gamma-Ray Bursts with Extended Emission}.
\newblock \apj 643 (2006) 266--275.

\bibitem{horvath10}
I.~{Horv{\'a}th}, et~al.
\newblock {Detailed Classification of Swift 's Gamma-ray Bursts}.
\newblock \apj 713 (2010) 552--557.

\bibitem{galama98}
T.~J. {Galama}, et~al.
\newblock {An unusual supernova in the error box of the {$\gamma$}-ray burst of
  25 April 1998}.
\newblock \nat 395 (1998) 670--672.

\bibitem{hjorth03}
J.~{Hjorth}, et~al.
\newblock {A very energetic supernova associated with the {$\gamma$}-ray burst
  of 29 March 2003}.
\newblock \nat 423 (2003) 847--850.

\bibitem{stanek03}
K.~Z. {Stanek}, et~al.
\newblock {Spectroscopic Discovery of the Supernova 2003dh Associated with GRB
  030329}.
\newblock \apjl 591 (2003) L17--L20.

\bibitem{campana06}
S.~{Campana}, et~al.
\newblock {The association of GRB 060218 with a supernova and the evolution of
  the shock wave}.
\newblock \nat 442 (2006) 1008--1010.

\bibitem{pian06}
E.~{Pian}, et~al.
\newblock {An optical supernova associated with the X-ray flash XRF 060218}.
\newblock \nat 442 (2006) 1011--1013.

\bibitem{fruchter06}
A.~S. {Fruchter}, et~al.
\newblock {Long {$\gamma$}-ray bursts and core-collapse supernovae have
  different environments}.
\newblock \nat 441 (2006) 463--468.

\bibitem{woosley93}
S.~E. {Woosley}.
\newblock {Gamma-ray bursts from stellar mass accretion disks around black
  holes}.
\newblock \apj 405 (1993) 273--277.

\bibitem{paczynski98}
B.~{Pacz\'ynski}.
\newblock {Are Gamma-Ray Bursts in Star-Forming Regions?}
\newblock \apjl 494 (1998) L45-L48.

\bibitem{gehrels05}
N.~{Gehrels}, et~al.
\newblock {A short {$\gamma$}-ray burst apparently associated with an
  elliptical galaxy at redshift z = 0.225}.
\newblock \nat 437 (2005) 851--854.

\bibitem{fox05}
D.~B. {Fox}, et~al.
\newblock {The afterglow of GRB 050709 and the nature of the short-hard
  {$\gamma$}-ray bursts}.
\newblock \nat 437 (2005) 845--850.

\bibitem{barthelmy05a}
S.~D. {Barthelmy}, et~al.
\newblock {An origin for short {$\gamma$}-ray bursts unassociated with current
  star formation}.
\newblock \nat 438 (2005) 994--996.

\bibitem{berger05}
E.~{Berger}, et~al.
\newblock {Afterglows, Redshifts, and Properties of Swift Gamma-Ray Bursts}.
\newblock \apj 634 (2005) 501--508.

\bibitem{paczynski86}
B.~{Pacz\'ynski}.
\newblock {Gamma-ray bursters at cosmological distances}.
\newblock \apjl 308 (1986) L43--L46.

\bibitem{eichler89}
D.~{Eichler}, et~al.
\newblock {Nucleosynthesis, neutrino bursts and gamma-rays from coalescing
  neutron stars}.
\newblock \nat 340 (1989) 126--128.

\bibitem{narayan92}
R.~{Narayan}, B.~{Paczynski}, T.~{Piran}.
\newblock {Gamma-ray bursts as the death throes of massive binary stars}.
\newblock \apjl 395 (1992) L83--L86.

\bibitem{gehrels06}
N.~{Gehrels}, et~al.
\newblock {A new {$\gamma$}-ray burst classification scheme from GRB060614}.
\newblock \nat 444 (2006) 1044--1046.

\bibitem{galyam06}
A.~{Gal-Yam}, et~al.
\newblock {A novel explosive process is required for the {$\gamma$}-ray burst
  GRB 060614}.
\newblock \nat 444 (2006) 1053--1055.

\bibitem{fynbo06}
J.~P.~U. {Fynbo}, et~al.
\newblock {No supernovae associated with two long-duration {$\gamma$}-ray
  bursts}.
\newblock \nat 444 (2006) 1047--1049.

\bibitem{dellavalle06}
M.~{Della Valle}, et~al.
\newblock {An enigmatic long-lasting {$\gamma$}-ray burst not accompanied by a
  bright supernova}.
\newblock \nat 444 (2006) 1050--1052.

\bibitem{zhang07b}
B.~{Zhang}, et~al.
\newblock {Making a Short Gamma-Ray Burst from a Long One: Implications for the
  Nature of GRB 060614}.
\newblock \apjl 655 (2007) L25--L28.

\bibitem{zhang09}
B.~{Zhang}, et~al.
\newblock {Discerning the Physical Origins of Cosmological Gamma-ray Bursts
  Based on Multiple Observational Criteria: The Cases of z = 6.7 GRB 080913, z
  = 8.2 GRB 090423, and Some Short/Hard GRBs}.
\newblock \apj 703 (2009) 1696--1724.

\bibitem{tanvir09}
N.~R. {Tanvir}, et~al.
\newblock {A {$\gamma$}-ray burst at a redshift of z\~{}8.2}.
\newblock \nat 461 (2009) 1254--1257.

\bibitem{salvaterra09}
R.~{Salvaterra}, et~al.
\newblock {GRB090423 at a redshift of z\~{}8.1}.
\newblock \nat 461 (2009) 1258--1260.

\bibitem{greiner09}
J.~{Greiner}, et~al.
\newblock {GRB 080913 at Redshift 6.7}.
\newblock \apj 693 (2009) 1610--1620.

\bibitem{levesque10}
E.~M. {Levesque}, et~al.
\newblock {GRB090426: the environment of a rest-frame 0.35-s gamma-ray burst at
  a redshift of 2.609}.
\newblock \mnras 401 (2010) 963--972.

\bibitem{antonelli10}
L.~A. {Antonelli}, et~al.
\newblock {GRB 090426: the farthest short gamma-ray burst?}
\newblock \aap 507 (2009) L45--L48.

\bibitem{xin10}
L.-P. {Xin}, et~al.
\newblock {Probing the nature of high-z short GRB 090426 with its early optical and X-ray afterglows}
\newblock \mnras 410 (2011) 27--32.


\bibitem{bloom08}
J.~S. {Bloom}, N.~R. {Butler}, D.~A. {Perley}.
\newblock {Gamma-ray Bursts, Classified Physically}.
\newblock In M.~{Galassi}, D.~{Palmer}, E.~{Fenimore}, editors, American
  Institute of Physics Conference Series, volume 1000 of American Institute of
  Physics Conference Series (2008)\hspace{0pt} 11--15.

\bibitem{donaghy06}
T.~Q. {Donaghy}, et~al.
\newblock {HETE-2 Localizations and Observations of Four Short Gamma-Ray
  Bursts: GRBs 010326B, 040802, 051211 and 060121}.
\newblock ArXiv Astrophysics e-prints: astro-ph/0605570  (2006).

\bibitem{kann10}
D. A. {Kann}, et~al.
\newblock {The Afterglows of Swift-era Gamma-ray Bursts. I. Comparing pre-Swift 
and Swift-era Long/Soft (Type II) GRB Optical Afterglows}.
\newblock \apj 720 (2010) 1513--1558.

\bibitem{kann08}
D. A. {Kann}, et~al.
\newblock {The Afterglows of Swift-era Gamma-Ray Bursts. II. Short/Hard (Type I) 
vs. Long/Soft (Type II) Optical Afterglows}.
\newblock \apj in press (2011) arXiv:0804.1959.

\bibitem{lv10}
H.-J. {L\"u}, E.-W. Liang, B.-B. Zhang, B. Zhang
\newblock {A New Classification Method for Gamma-Ray Bursts}.
\newblock \apj 725 (2010) 1965--1970.


\bibitem{starling10}
R. L. C. {Starling}, et~al.
\newblock {Discovery of the nearby long, soft GRB 100316D with an associated supernova}.
\newblock \mnras 411 (2011) 2792--2803.


\bibitem{savaglio09}
S.~{Savaglio}, K.~{Glazebrook}, D.~{Le Borgne}.
\newblock {The Galaxy Population Hosting Gamma-Ray Bursts}.
\newblock \apj 691 (2009) 182-211.

\bibitem{woosleyheger06}
S.~E. {Woosley}, A.~{Heger}.
\newblock {The Progenitor Stars of Gamma-Ray Bursts}.
\newblock \apj 637 (2006) 914--921.

\bibitem{fryer99}
C.~L. {Fryer}, S.~E. {Woosley}, D.~H. {Hartmann}.
\newblock {Formation Rates of Black Hole Accretion Disk Gamma-Ray Bursts}.
\newblock \apj 526 (1999) 152--177.

\bibitem{yoon05}
{S.-C.} {Yoon}, N.~{Langer}.
\newblock {Evolution of rapidly rotating metal-poor massive stars towards
  gamma-ray bursts}.
\newblock \aap 443 (2005) 643--648.

\bibitem{vietri98}
M.~{Vietri}, L.~{Stella}.
\newblock {A Gamma-Ray Burst Model with Small Baryon Contamination}.
\newblock \apjl 507 (1998) L45--L48.

\bibitem{soderberg06}
A.~M. {Soderberg}, et~al.
\newblock {Relativistic ejecta from X-ray flash XRF 060218 and the rate of
  cosmic explosions}.
\newblock \nat 442 (2006) 1014--1017.

\bibitem{mazzali06}
P.~A. {Mazzali}, et~al.
\newblock {A neutron-star-driven X-ray flash associated with supernova SN
  2006aj}.
\newblock \nat 442 (2006) 1018--1020.

\bibitem{liang07}
E.~{Liang}, et~al.
\newblock {Low-Luminosity Gamma-Ray Bursts as a Unique Population: Luminosity
  Function, Local Rate, and Beaming Factor}.
\newblock \apj 662 (2007) 1111--1118.

\bibitem{holland10}
S.~T. {Holland}, et~al.
\newblock {GRB 090417B and its Host Galaxy: A Step Toward an Understanding of
  Optically Dark Gamma-ray Bursts}.
\newblock \apj 717 (2010) 223--234.

\bibitem{wolf07}
C.~{Wolf}, P.~{Podsiadlowski}.
\newblock {The metallicity dependence of the long-duration gamma-ray burst rate
  from host galaxy luminosities}.
\newblock \mnras 375 (2007) 1049--1058.

\bibitem{li08}
L.-X. {Li}.
\newblock {Star formation history up to z = 7.4: implications for gamma-ray
  bursts and cosmic metallicity evolution}.
\newblock \mnras 388 (2008) 1487--1500.

\bibitem{niino10}
Y. {Niino}, et~al.
\newblock {Luminosity Distribution of Gamma-Ray Burst Host Galaxies at redshift 
$z=1$ in Cosmological Smoothed Particle Hydrodinamic Simulations: Implications 
for the Metallicity Dependence of GRBs}.
\newblock \apj 726 (2011) 88.

\bibitem{bloom06}
J.~S. {Bloom}, et~al.
\newblock {Closing in on a Short-Hard Burst Progenitor: Constraints from
  Early-Time Optical Imaging and Spectroscopy of a Possible Host Galaxy of GRB
  050509b}.
\newblock \apj 638 (2006) 354--368.

\bibitem{fong10}
W.~{Fong}, E.~{Berger}, D.~B. {Fox}.
\newblock {Hubble Space Telescope Observations of Short Gamma-Ray Burst Host
  Galaxies: Morphologies, Offsets, and Local Environments}.
\newblock \apj 708 (2010) 9--25.

\bibitem{hjorth05}
J.~{Hjorth}, et~al.
\newblock {The optical afterglow of the short {$\gamma$}-ray burst GRB 050709}.
\newblock \nat 437 (2005) 859--861.

\bibitem{rosswog03}
S.~{Rosswog}, E.~{Ramirez-Ruiz}, M.~B. {Davies}.
\newblock {High-resolution calculations of merging neutron stars - III.
  Gamma-ray bursts}.
\newblock \mnras 345 (2003) 1077--1090.

\bibitem{aloy05}
M.~A. {Aloy}, H.-T. {Janka}, E.~{M{\"u}ller}.
\newblock {Relativistic outflows from remnants of compact object mergers and
  their viability for short gamma-ray bursts}.
\newblock \aap 436 (2005) 273--311.

\bibitem{rezzolla11}
L.~{Rezzolla} et al.
\newblock {The missing link: merging neutron stars naturally produce 
jet-like structures and can power short Gamma-Ray Bursts}.
\newblock preprint (arXiv:1101.4298).

\bibitem{paczynski91}
B.~{Pacz\'ynski}.
\newblock {Cosmological gamma-ray bursts}.
\newblock Acta Astronomica 41 (1991) 257--267.

\bibitem{taylor89}
J.~H. {Taylor}, J.~M. {Weisberg}.
\newblock {Further experimental tests of relativistic gravity using the binary
  pulsar PSR 1913 + 16}.
\newblock \apj 345 (1989) 434--450.

\bibitem{kramer08}
M.~{Kramer}, I.~H. {Stairs}.
\newblock {The Double Pulsar}.
\newblock \araa 46 (2008) 541--572.

\bibitem{nakar07}
E.~{Nakar}.
\newblock {Short-hard gamma-ray bursts}.
\newblock \physrep 442 (2007) 166--236.

\bibitem{lee07}
W.~H. {Lee}, E.~{Ramirez-Ruiz}.
\newblock {The progenitors of short gamma-ray bursts}.
\newblock New Journal of Physics 9 (2007) 17--+.

\bibitem{berger07}
E.~{Berger}, et~al.
\newblock {A New Population of High-Redshift Short-Duration Gamma-Ray Bursts}.
\newblock \apj 664 (2007) 1000--1010.

\bibitem{deugartepostigo06}
A.~{de Ugarte Postigo}, et~al.
\newblock {GRB 060121: Implications of a Short-/Intermediate-Duration
  {$\gamma$}-Ray Burst at High Redshift}.
\newblock \apjl 648 (2006) L83--L87.

\bibitem{virgili10}
F.~J. {Virgili}, B. Zhang, P. O'Brien, E. Troja.
\newblock {Are all short-hard gamma-ray bursts produced from mergers of 
compact stellar objects?}
\newblock \apj 727 (2011) 109.

\bibitem{nakar06}
E.~{Nakar}, A.~{Gal-Yam}, D.~B. {Fox}.
\newblock {The Local Rate and the Progenitor Lifetimes of Short-Hard Gamma-Ray
  Bursts: Synthesis and Predictions for the Laser Interferometer
  Gravitational-Wave Observatory}.
\newblock \apj 650 (2006) 281--290.

\bibitem{guetta06}
D.~{Guetta}, T.~{Piran}.
\newblock {The BATSE-Swift luminosity and redshift distributions of
  short-duration GRBs}.
\newblock \aap 453 (2006) 823--828.

\bibitem{lee10}
W.~H. {Lee}, E. {Ramirez-Ruiz}, G. {van de Ven}.
\newblock {Short Gamma-ray Bursts from Dynamically Assembled Compact 
Binaries in Globular Clusters: Pathways, Rates, Hydrodynamics, and 
Cosmological Setting}.
\newblock \apj 720 (2010) 953--975.

\bibitem{rosswog11}
S.~{Rosswog}
\newblock invited talk at ``Prompt GRB 2011'', Raleigh. 

\bibitem{cutler02}
C.~{Cutler}, K.~S. {Thorne}.
\newblock {An Overview of Gravitational-Wave Sources}.
\newblock ArXiv:gr-qc/0204090 (2002).

\bibitem{dermer06}
C.~D. {Dermer}, A.~{Atoyan}.
\newblock {Collapse of Neutron Stars to Black Holes in Binary Systems: A Model
  for Short Gamma-Ray Bursts}.
\newblock \apjl 643 (2006) L13--L16.

\bibitem{palmer05}
D.~M. {Palmer}, et~al.
\newblock {A giant {$\gamma$}-ray flare from the magnetar SGR 1806 - 20}.
\newblock \nat 434 (2005) 1107--1109.

\bibitem{hurley05}
K.~{Hurley}, et~al.
\newblock {An exceptionally bright flare from SGR 1806-20 and the origins of
  short-duration {$\gamma$}-ray bursts}.
\newblock \nat 434 (2005) 1098--1103.

\bibitem{tanvir05}
N.~R. {Tanvir}, et~al.
\newblock {An origin in the local Universe for some short {$\gamma$}-ray
  bursts}.
\newblock \nat 438 (2005) 991--993.

\bibitem{nakar06b}
E.~{Nakar}, et~al.
\newblock {The Distances of Short-Hard Gamma-Ray Bursts and the Soft Gamma-Ray
  Repeater Connection}.
\newblock \apj 640 (2006) 849--853.

\bibitem{fishman95}
G.~J. {Fishman}, C.~A. {Meegan}.
\newblock {Gamma-Ray Bursts}.
\newblock \araa 33 (1995) 415--458.

\bibitem{zhangmeszaros04}
B.~{Zhang}, P.~{M{\'e}sz{\'a}ros}.
\newblock {Gamma-Ray Bursts: progress, problems \& prospects}.
\newblock International Journal of Modern Physics A 19 (2004) 2385--2472.

\bibitem{piran99}
T.~{Piran}.
\newblock {Gamma-ray bursts and the fireball model}.
\newblock \physrep 314 (1999) 575--667.

\bibitem{lithwick01}
Y.~{Lithwick}, R.~{Sari}.
\newblock {Lower Limits on Lorentz Factors in Gamma-Ray Bursts}.
\newblock \apj 555 (2001) 540--545.

\bibitem{liang10}
E.-W.~{Liang}, et~al.
\newblock {Constraining GRB Initial Lorentz Factor with the Afterglow Onset 
Feature and Discovery of a Tight $\Gamma_0-E_{iso}$ Correlation}.
\newblock \apj 725 (2010) 2209--2224.

\bibitem{abdo09a}
A.~A. {Abdo}, et~al.
\newblock {Fermi Observations of High-Energy Gamma-Ray Emission from GRB
  080916C}.
\newblock Science 323 (2009) 1688--1693.

\bibitem{abdo09b}
A.~A. {Abdo}, et~al.
\newblock {A limit on the variation of the speed of light arising from quantum
  gravity effects}.
\newblock \nat 462 (2009) 331--334.


\bibitem{abdo09c}
A.~A. {Abdo}, et~al.
\newblock {Fermi Observations of GRB 090902B: A Distinct Spectral Component in
  the Prompt and Delayed Emission}.
\newblock \apjl 706 (2009) L138--L144.


\bibitem{frail01}
D.~A. {Frail}, et~al.
\newblock {Beaming in Gamma-Ray Bursts: Evidence for a Standard Energy
  Reservoir}.
\newblock \apjl 562 (2001) L55--L58.

\bibitem{bloom03}
J.~S. {Bloom}, D.~A. {Frail}, S.~R. {Kulkarni}.
\newblock {Gamma-Ray Burst Energetics and the Gamma-Ray Burst Hubble Diagram:
  Promises and Limitations}.
\newblock \apj 594 (2003) 674--683.

\bibitem{liang08}
E.-W. {Liang}, et~al.
\newblock {A Comprehensive Analysis of Swift XRT Data. III. Jet Break
  Candidates in X-Ray and Optical Afterglow Light Curves}.
\newblock \apj 675 (2008) 528--552.

\bibitem{racusin09}
J.~L. {Racusin}, et~al.
\newblock {Jet breaks and Energetics of Swift GRB X-ray Afterglows}.
\newblock \apj 698 (2009) 43--74.

\bibitem{beloborodov98}
A.~M. {Beloborodov}, B.~E. {Stern}, R.~{Svensson}.
\newblock {Self-Similar Temporal Behavior of Gamma-Ray Bursts}.
\newblock \apjl 508 (1998) L25--L27.

\bibitem{norris02}
J.~P. {Norris}.
\newblock {Implications of the Lag-Luminosity Relationship for Unified
  Gamma-Ray Burst Paradigms}.
\newblock \apj 579 (2002) 386--403.

\bibitem{macfadyen99}
A.~I. {MacFadyen}, S.~E. {Woosley}.
\newblock {Collapsars: Gamma-Ray Bursts and Explosions in ``Failed
  Supernovae''}.
\newblock \apj 524 (1999) 262--289.

\bibitem{proga03}
D.~{Proga}, et~al.
\newblock {Axisymmetric Magnetohydrodynamic Simulations of the Collapsar Model
  for Gamma-Ray Bursts}.
\newblock \apjl 599 (2003) L5--L8.

\bibitem{zhangw03}
W.~{Zhang}, S.~E. {Woosley}, A.~I. {MacFadyen}.
\newblock {Relativistic Jets in Collapsars}.
\newblock \apj 586 (2003) 356--371.

\bibitem{morsony10}
B.~J. {Morsony}, D.~{Lazzati}, M.~C. {Begelman}.
\newblock {The Origin and Propagation of Variability in the 
Outflows of Long-duration Gamma-ray Bursts}.
\newblock \apj 723 (2010) 267--276.

\bibitem{blandford77}
R.~D. {Blandford}, R.~L. {Znajek}.
\newblock {Electromagnetic extraction of energy from Kerr black holes}.
\newblock \mnras 179 (1977) 433--456.

\bibitem{meszarosrees97b}
P.~{M\'esz\'aros}, M.~J. {Rees}.
\newblock {Poynting Jets from Black Holes and Cosmological Gamma-Ray Bursts}.
\newblock \apjl 482 (1997) L29-L32.

\bibitem{lee00}
H.~K. {Lee}, R.~A.~M.~J. {Wijers}, G.~E. {Brown}.
\newblock {The Blandford-Znajek process as a central engine for a gamma-ray
  burst}.
\newblock \physrep 325 (2000) 83--114.

\bibitem{li00}
{L.-X.} {Li}.
\newblock {Extracting Energy from a Black Hole through Its Disk}.
\newblock \apjl 533 (2000) L115--L118.

\bibitem{mckinney05}
J.~C. {McKinney}.
\newblock {Total and Jet Blandford-Znajek Power in the Presence of an Accretion
  Disk}.
\newblock \apjl 630 (2005) L5--L8.

\bibitem{proga03b}
D.~{Proga}, M.~C. {Begelman}.
\newblock {Accretion of Low Angular Momentum Material onto Black Holes:
  Two-dimensional Magnetohydrodynamic Case}.
\newblock \apj 592 (2003) 767--781.

\bibitem{lei09}
W.~H. {Lei}, et~al.
\newblock {Magnetically Torqued Neutrino-dominated Accretion Flows for
  Gamma-ray Bursts}.
\newblock \apj 700 (2009) 1970--1976.

\bibitem{usov92}
V.~V. {Usov}.
\newblock {Millisecond pulsars with extremely strong magnetic fields as a
  cosmological source of gamma-ray bursts}.
\newblock \nat 357 (1992) 472--474.

\bibitem{zhangdai09}
D.~{Zhang}, Z.~G. {Dai}.
\newblock {Hyperaccreting Neutron Star Disks and Neutrino Annihilation}.
\newblock \apj 703 (2009) 461--478.

\bibitem{janka95}
{H.-T.} {Janka}, E.~{Mueller}.
\newblock {The First Second of a Type II Supernova: Convection, Accretion, and
  Shock Propagation}.
\newblock \apjl 448 (1995) L109--L113.

\bibitem{qian96}
{Y.-Z.} {Qian}, S.~E. {Woosley}.
\newblock {Nucleosynthesis in Neutrino-driven Winds. I. The Physical Condition}.
\newblock \apj 471 (1996) 331--351.

\bibitem{thompson04}
T.~A. {Thompson}, P.~{Chang}, E.~{Quataert}.
\newblock {Magnetar Spin-Down, Hyperenergetic Supernovae, and Gamma-Ray
  Bursts}.
\newblock \apj 611 (2004) 380--393.

\bibitem{metzger07}
B.~D. {Metzger}, T.~A. {Thompson}, E.~{Quataert}.
\newblock {Proto-Neutron Star Winds with Magnetic Fields and Rotation}.
\newblock \apj 659 (2007) 561--579.

\bibitem{metzger11}
B.~D. {Metzger} et al.
\newblock {The Proto-Magnetar Model for Gamma-Ray Bursts}.
\newblock \mnras submitted (arXiv:1012.0001).

\bibitem{wheeler00}
J.~C. {Wheeler}, et~al.
\newblock {Asymmetric Supernovae, Pulsars, Magnetars, and Gamma-Ray Bursts}.
\newblock \apj 537 (2000) 810--823.

\bibitem{lyutikov03}
M.~{Lyutikov}, R.~{Blandford}.
\newblock {Gamma Ray Bursts as Electromagnetic Outflows}.
\newblock ArXiv Astrophysics e-prints astro-ph/0312347 (2003).

\bibitem{witten84}
E.~{Witten}.
\newblock {Cosmic separation of phases}.
\newblock \prd 30 (1984) 272--285.

\bibitem{alcock86}
C.~{Alcock}, E.~{Farhi}, A.~{Olinto}.
\newblock {Strange stars}.
\newblock \apj 310 (1986) 261--272.

\bibitem{chengdai96}
K.~S. {Cheng}, Z.~G. {Dai}.
\newblock {Conversion of Neutron Stars to Strange Stars as a Possible Origin of
  {$\gamma$}-Ray Bursts}.
\newblock Physical Review Letters 77 (1996) 1210--1213.

\bibitem{dailu98a}
Z.~G. {Dai}, T.~{Lu}.
\newblock {{$\gamma$}-Ray Bursts and Afterglows from Rotating Strange Stars and
  Neutron Stars}.
\newblock Physical Review Letters 81 (1998) 4301--4304.

\bibitem{ouyed05}
R.~{Ouyed}, R.~{Rapp}, C.~{Vogt}.
\newblock {Fireballs from Quark Stars in the Color-Flavor Locked Phase:
  Application to Gamma-Ray Bursts}.
\newblock \apj 632 (2005) 1001--1007.

\bibitem{paczynski05}
B.~{Paczy{\'n}ski}, P.~{Haensel}.
\newblock {Gamma-ray bursts from quark stars}.
\newblock \mnras 362 (2005) L4--L7.

\bibitem{xuliang09}
R.~{Xu}, E.~{Liang}.
\newblock {X-ray flares of {$\gamma$}-ray bursts: Quakes of solid quark stars?}
\newblock Science in China G: Physics and Astronomy 52 (2009) 315--320.

\bibitem{iwamoto98}
K.~{Iwamoto}, et~al.
\newblock {A hypernova model for the supernova associated with the
  {$\gamma$}-ray burst of 25 April 1998}.
\newblock \nat 395 (1998) 672--674.

\bibitem{mazzali03}
P.~A. {Mazzali}, et~al.
\newblock {The Type Ic Hypernova SN 2003dh/GRB 030329}.
\newblock \apjl 599 (2003) L95--L98.

\bibitem{dailu98b}
Z.~G. {Dai}, T.~{Lu}.
\newblock {Gamma-ray burst afterglows and evolution of postburst fireballs with
  energy injection from strongly magnetic millisecond pulsars}.
\newblock \aap 333 (1998) L87--L90.

\bibitem{zhangmeszaros01a}
B.~{Zhang}, P.~{M{\'e}sz{\'a}ros}.
\newblock {Gamma-Ray Burst Afterglow with Continuous Energy Injection:
  Signature of a Highly Magnetized Millisecond Pulsar}.
\newblock \apjl 552 (2001) L35--L38.

\bibitem{zhang06}
B.~{Zhang}, et~al.
\newblock {Physical Processes Shaping Gamma-Ray Burst X-Ray Afterglow Light
  Curves: Theoretical Implications from the Swift X-Ray Telescope
  Observations}.
\newblock \apj 642 (2006) 354--370.

\bibitem{nousek06}
J.~A. {Nousek}, et~al.
\newblock {Evidence for a Canonical Gamma-Ray Burst Afterglow Light Curve in
  the Swift XRT Data}.
\newblock \apj 642 (2006) 389--400.

\bibitem{obrien06}
P.~T. {O'Brien}, et~al.
\newblock {The Early X-Ray Emission from GRBs}.
\newblock \apj 647 (2006) 1213--1237.

\bibitem{troja07}
E.~{Troja}, et~al.
\newblock {Swift Observations of GRB 070110: An Extraordinary X-Ray Afterglow
  Powered by the Central Engine}.
\newblock \apj 665 (2007) 599--607.

\bibitem{liang07b}
E.-W. {Liang}, B.-B. {Zhang}, B.~{Zhang}.
\newblock {A Comprehensive Analysis of Swift XRT Data. II. Diverse Physical
  Origins of the Shallow Decay Segment}.
\newblock \apj 670 (2007) 565--583.

\bibitem{lyons10}
N.~{Lyons}, et~al.
\newblock {Can X-ray emission powered by a spinning-down magnetar explain some
  gamma-ray burst light-curve features?}
\newblock \mnras 402 (2010) 705--712.

\bibitem{novak10}
J.~{Novak}.
\newblock {Numerical simulations of GRB engines}.
\newblock this volume.


\bibitem{derishev99}
E.~V. {Derishev}, V.~V. {Kocharovsky}, V.~V. {Kocharovsky}.
\newblock {The Neutron Component in Fireballs of Gamma-Ray Bursts: Dynamics and
  Observable Imprints}.
\newblock \apj 521 (1999) 640--649.

\bibitem{beloborodov03}
A.~M. {Beloborodov}.
\newblock {Nuclear Composition of Gamma-Ray Burst Fireballs}.
\newblock \apj 588 (2003) 931--944.

\bibitem{beloborodov03b}
A.~M. {Beloborodov}.
\newblock {Neutron-fed Afterglows of Gamma-Ray Bursts}.
\newblock \apjl 585 (2003) L19--L22.

\bibitem{fan05b}
Y.~Z. {Fan}, B.~{Zhang}, D.~M. {Wei}.
\newblock {Early Optical Afterglow Light Curves of Neutron-fed Gamma-Ray
  Bursts}.
\newblock \apj 628 (2005) 298--314.

\bibitem{goodman86}
J.~{Goodman}.
\newblock {Are gamma-ray bursts optically thick?}
\newblock \apjl 308 (1986) L47--L50.

\bibitem{shemi90}
A.~{Shemi}, T.~{Piran}.
\newblock {The appearance of cosmic fireballs}.
\newblock \apjl 365 (1990) L55--L58.

\bibitem{rees92}
M.~J. {Rees}, P.~{M\'esz\'aros}.
\newblock {Relativistic fireballs - Energy conversion and time-scales}.
\newblock \mnras 258 (1992) 41P--43P.

\bibitem{meszarosrees93}
P.~{M\'esz\'aros}, M.~J. {Rees}.
\newblock {Relativistic fireballs and their impact on external matter - Models
  for cosmological gamma-ray bursts}.
\newblock \apj 405 (1993) 278--284.

\bibitem{rees94}
M.~J. {Rees}, P.~{M\'esz\'aros}.
\newblock {Unsteady outflow models for cosmological gamma-ray bursts}.
\newblock \apjl 430 (1994) L93--L96.

\bibitem{drenkhahn02}
G.~{Drenkhahn}, H.~C. {Spruit}.
\newblock {Efficient acceleration and radiation in Poynting flux powered GRB
  outflows}.
\newblock \aap 391 (2002) 1141--1153.

\bibitem{vlahakis03}
N.~{Vlahakis}, A.~{K{\"o}nigl}.
\newblock {Relativistic Magnetohydrodynamics with Application to Gamma-Ray
  Burst Outflows. I. Theory and Semianalytic Trans-Alfv{\'e}nic Solutions}.
\newblock \apj 596 (2003) 1080--1103.

\bibitem{komissarov09}
S.~S. {Komissarov}, et~al.
\newblock {Magnetic acceleration of ultrarelativistic jets in gamma-ray burst
  sources}.
\newblock \mnras 394 (2009) 1182--1212.

\bibitem{tchekhovskoy09}
A.~{Tchekhovskoy}, J.~C. {McKinney}, R.~{Narayan}.
\newblock {Efficiency of Magnetic to Kinetic Energy Conversion in a Monopole
  Magnetosphere}.
\newblock \apj 699 (2009) 1789--1808.

\bibitem{zhangyan10}
B.~{Zhang}, H.~{Yan}.
\newblock {The Internal-Collision-Induced Magnetic Reconnection
and Turbulence (ICMART) Model of Gamma-Ray Bursts}.
\newblock \apj 726 (2011) 90.

\bibitem{waxman03}
E.~{Waxman}.
\newblock {Astronomy: New direction for {$\gamma$}-rays}.
\newblock \nat 423 (2003) 388--389.

\bibitem{lyutikov03b}
M.~{Lyutikov}, V.~I. {Pariev}, R.~D. {Blandford}.
\newblock {Polarization of Prompt Gamma-Ray Burst Emission: Evidence for
  Electromagnetically Dominated Outflow}.
\newblock \apj 597 (2003) 998--1009.

\bibitem{granot03}
J.~{Granot}.
\newblock {The Most Probable Cause for the High Gamma-Ray Polarization in GRB
  021206}.
\newblock \apjl 596 (2003) L17--L21.

\bibitem{toma09b}
K.~{Toma}, et~al.
\newblock {Statistical Properties of Gamma-Ray Burst Polarization}.
\newblock \apj 698 (2009) 1042--1053.

\bibitem{coburn03}
W.~{Coburn}, S.~E. {Boggs}.
\newblock {Polarization of the prompt {$\gamma$}-ray emission from the
  {$\gamma$}-ray burst of 6 December 2002}.
\newblock \nat 423 (2003) 415--417.

\bibitem{willis05}
D.~R. {Willis}, et~al.
\newblock {Evidence of polarisation in the prompt gamma-ray emission from GRB
  930131 and GRB 960924}.
\newblock \aap 439 (2005) 245--253.

\bibitem{rutledge04}
R.~E. {Rutledge}, D.~B. {Fox}.
\newblock {Re-analysis of polarization in the {$\gamma$}-ray flux of GRB
  021206}.
\newblock \mnras 350 (2004) 1288--1300.

\bibitem{zhangpeer09}
B.~{Zhang}, A.~{Pe'er}.
\newblock {Evidence of an Initially Magnetically Dominated Outflow in GRB
  080916C}.
\newblock \apjl 700 (2009) L65--L68.

\bibitem{fan10}
{Y.-Z.} {Fan}.
\newblock {The spectrum of {$\gamma$}-ray burst: a clue}.
\newblock \mnras 403 (2010) 483--490.

\bibitem{zhang10}
B.-B.~{Zhang} et~al.
\newblock {A Comprehensive Analysis of Fermi Gamma-Ray Burst Data. I. Spectral 
Components and Their Possible Physical Origins of LAT/GBM GRBs}.
\newblock \apj 730 (2011) 141.

\bibitem{beloborodov10}
A.~M. {Beloborodov}.
\newblock {Collisional mechanism for gamma-ray burst emission}.
\newblock \mnras 407 (2010) 1033--1047.

\bibitem{lazzati10}
D.~{Lazzati}, M.~C. {Begelman}
\newblock {Non-Thermal emission from the photospheres of Gamma-Ray Burst outflows. I: High frequency tails}.
\newblock \apj 725 (2010) 1137--1145.

\bibitem{ackermann10}
M.~{Ackermann}, et~al.
\newblock {Fermi Observations of GRB 090510: A Short-Hard Gamma-ray Burst with
  an Additional, Hard Power-law Component from 10 keV TO GeV Energies}.
\newblock \apj 716 (2010) 1178--1190.

\bibitem{ryde10}
F.~{Ryde}, et~al.
\newblock {Identification and Properties of the Photospheric Emission in
  GRB090902B}.
\newblock \apjl 709 (2010) L172--L177.

\bibitem{peer10}
A.~{Pe'er} et~al.
\newblock {The Connection Between Thermal and Non-Thermal Emission in Gamma-ray 
Bursts: General Considerations and GRB090902B as a Case Study}.
\newblock \apj submitted (2010) arXiv:1007.2228.


\bibitem{fan04b}
Y.~Z. {Fan}, D.~M. {Wei}, C.~F. {Wang}.
\newblock {The very early afterglow powered by ultra-relativistic mildly
  magnetized outflows}.
\newblock \aap 424 (2004) 477--484.

\bibitem{zhangkobayashi05}
B.~{Zhang}, S.~{Kobayashi}.
\newblock {Gamma-Ray Burst Early Afterglows: Reverse Shock Emission from an
  Arbitrarily Magnetized Ejecta}.
\newblock \apj 628 (2005) 315--334.

\bibitem{mimica09}
P.~{Mimica}, D.~{Giannios}, M.~A. {Aloy}.
\newblock {Deceleration of arbitrarily magnetized GRB ejecta: the complete
  evolution}.
\newblock \aap 494 (2009) 879--890.

\bibitem{mizuno09}
Y.~{Mizuno}, et~al.
\newblock {Magnetohydrodynamic Effects in Propagating Relativistic Jets:
  Reverse Shock and Magnetic Acceleration}.
\newblock \apjl 690 (2009) L47--L51.

\bibitem{fan02}
Y.-Z. {Fan}, et~al.
\newblock {Optical Flash of GRB 990123: Constraints on the Physical Parameters
  of the Reverse Shock}.
\newblock Chinese Journal of Astronomy and Astrophysics 2 (2002) 449--453.

\bibitem{zhang03}
B.~{Zhang}, S.~{Kobayashi}, P.~{M{\'e}sz{\'a}ros}.
\newblock {Gamma-Ray Burst Early Optical Afterglows: Implications for the
  Initial Lorentz Factor and the Central Engine}.
\newblock \apj 595 (2003) 950--954.

\bibitem{kumar03}
P.~{Kumar}, A.~{Panaitescu}.
\newblock {A unified treatment of the gamma-ray burst 021211 and its
  afterglow}.
\newblock \mnras 346 (2003) 905--914.

\bibitem{gomboc08}
A.~{Gomboc}, et~al.
\newblock {Multiwavelength Analysis of the Intriguing GRB 061126: The Reverse
  Shock Scenario and Magnetization}.
\newblock \apj 687 (2008) 443--455.

\bibitem{steele09}
I.~A. {Steele}, et~al.
\newblock {Ten per cent polarized optical emission from GRB 090102.}
\newblock \nat 462 (2009) 767--769.

\bibitem{mckinney11}
J.~C. McKinney, D.~A. Uzdensky.
\newblock {A Reconnection Switch to Trigger Gamma-Ray Burst Jet Dissipation}
\newblock \mnras, submitted (arXiv:1011.1904).

\bibitem{tagliaferri05}
G.~{Tagliaferri}, et~al.
\newblock {An unexpectedly rapid decline in the X-ray afterglow emission of
  long {$\gamma$}-ray bursts}.
\newblock \nat 436 (2005) 985--988.

\bibitem{barthelmy05b}
S.~D. {Barthelmy}, et~al.
\newblock {Discovery of an Afterglow Extension of the Prompt Phase of Two
  Gamma-Ray Bursts Observed by Swift}.
\newblock \apjl 635 (2005) L133--L136.

\bibitem{zhangbb07}
B.-B. {Zhang}, E.-W. {Liang}, B.~{Zhang}.
\newblock {A Comprehensive Analysis of Swift XRT Data. I. Apparent Spectral
  Evolution of Gamma-Ray Burst X-Ray Tails}.
\newblock \apj 666 (2007) 1002--1011.

\bibitem{kumar00}
P.~{Kumar}, A.~{Panaitescu}.
\newblock {Afterglow Emission from Naked Gamma-Ray Bursts}.
\newblock \apjl 541 (2000) L51--L54.

\bibitem{liang06}
E.~W. {Liang}, et~al.
\newblock {Testing the Curvature Effect and Internal Origin of Gamma-Ray Burst
  Prompt Emissions and X-Ray Flares with Swift Data}.
\newblock \apj 646 (2006) 351--357.

\bibitem{zhangbb09}
B.-B. {Zhang}, et~al.
\newblock {Curvature Effect of a Non-Power Spectrum and Spectral Evolution of
  GRB X-Ray Tails}.
\newblock \apjl 690 (2009) L10--L13.

\bibitem{lyutikov06}
M.~{Lyutikov}.
\newblock {Did Swift measure gamma-ray burst prompt emission radii?}
\newblock \mnras 369 (2006) L5--L8.

\bibitem{kumar06}
P.~{Kumar}, et~al.
\newblock {A unified picture for gamma-ray burst prompt and X-ray afterglow
  emissions}.
\newblock \mnras 367 (2006) L52--L56.

\bibitem{vestrand05}
W.~T. {Vestrand}, et~al.
\newblock {A link between prompt optical and prompt {$\gamma$}-ray emission in
  {$\gamma$}-ray bursts}.
\newblock \nat 435 (2005) 178--180.

\bibitem{racusin08}
J.~L. {Racusin}, et~al.
\newblock {Broadband observations of the naked-eye {$\gamma$}-ray burst
  GRB080319B}.
\newblock \nat 455 (2008) 183--188.

\bibitem{kumarpanaitescu08}
P.~{Kumar}, A.~{Panaitescu}.
\newblock {What did we learn from gamma-ray burst 080319B?}
\newblock \mnras 391 (2008) L19--L23.

\bibitem{shen09}
{R.-F.} {Shen}, B.~{Zhang}.
\newblock {Prompt optical emission and synchrotron self-absorption constraints
  on emission site of GRBs}.
\newblock \mnras 398 (2009) 1936--1950.

\bibitem{gupta08}
N.~{Gupta}, B.~{Zhang}.
\newblock {Diagnosing the site of gamma-ray burst prompt emission with spectral
  cut-off energy}.
\newblock \mnras 384 (2008) L11--L15.

\bibitem{kumar07}
P.~{Kumar}, et~al.
\newblock {The nature of the outflow in gamma-ray bursts}.
\newblock \mnras 376 (2007) L57--L61.

\bibitem{kumarmcmahon08}
P.~{Kumar}, E.~{McMahon}.
\newblock {A general scheme for modelling {$\gamma$}-ray burst prompt
  emission}.
\newblock \mnras 384 (2008) 33--63.

\bibitem{fan09}
{Y.-Z.} {Fan}, B.~{Zhang}, {D.-M.} {Wei}.
\newblock {Naked-eye optical flash from gamma-ray burst 080319B: Tracing the
  decaying neutrons in the outflow}.
\newblock \prd 79 (2009) 2 021301.

\bibitem{zou09}
{Y.-C.} {Zou}, T.~{Piran}, R.~{Sari}.
\newblock {Clues from the Prompt Emission of GRB 080319B}.
\newblock \apjl 692 (2009) L92--L95.

\bibitem{resmi10}
L.~{Resmi}, B.~{Zhang}.
\newblock {Gamma Ray Burst Prompt Emission Variability in Synchrotron and 
Synchrotron Self-Compton Lightcurves}.
\newblock \mnras (2010) submitted.

\bibitem{waxman95}
E.~{Waxman}.
\newblock {Cosmological Gamma-Ray Bursts and the Highest Energy Cosmic Rays}.
\newblock Physical Review Letters 75 (1995) 386--389.

\bibitem{vietri95}
M.~{Vietri}.
\newblock {The Acceleration of Ultra--High-Energy Cosmic Rays in Gamma-Ray
  Bursts}.
\newblock \apj 453 (1995) 883--889.

\bibitem{waxman97}
E.~{Waxman}, J.~{Bahcall}.
\newblock {High Energy Neutrinos from Cosmological Gamma-Ray Burst Fireballs}.
\newblock Physical Review Letters 78 (1997) 2292--2295.

\bibitem{meszaroswaxman01}
P.~{M{\'e}sz{\'a}ros}, E.~{Waxman}.
\newblock {TeV Neutrinos from Successful and Choked Gamma-Ray Bursts}.
\newblock Physical Review Letters 87 (2001) 17 171102.

\bibitem{razzaque04b}
S.~{Razzaque}, P.~{M{\'e}sz{\'a}ros}, E.~{Waxman}.
\newblock {Neutrino signatures of the supernova: Gamma ray burst relationship}.
\newblock \prd 69 (2004) 2 023001.

\bibitem{paczynski94}
B.~{Pacz\'ynski}, G.~{Xu}.
\newblock {Neutrino bursts from gamma-ray bursts}.
\newblock \apj 427 (1994) 708--713.

\bibitem{meszaros93}
P.~{M\'esz\'aros}, P.~{Laguna}, M.~J. {Rees}.
\newblock {Gasdynamics of relativistically expanding gamma-ray burst sources -
  Kinematics, energetics, magnetic fields, and efficiency}.
\newblock \apj 415 (1993) 181--190.

\bibitem{meszarosrees97}
P.~{M\'esz\'aros}, M.~J. {Rees}.
\newblock {Optical and Long-Wavelength Afterglow from Gamma-Ray Bursts}.
\newblock \apj 476 (1997) 232.

\bibitem{saripiran99}
R.~{Sari}, T.~{Piran}.
\newblock {GRB 990123: The Optical Flash and the Fireball Model}.
\newblock \apjl 517 (1999) L109--L112.

\bibitem{kobayashi00}
S.~{Kobayashi}.
\newblock {Light Curves of Gamma-Ray Burst Optical Flashes}.
\newblock \apj 545 (2000) 807--812.

\bibitem{sari95}
R.~{Sari}, T.~{Piran}.
\newblock {Hydrodynamic Timescales and Temporal Structure of Gamma-Ray Bursts}.
\newblock \apjl 455 (1995) L143--L146.

\bibitem{blandford76}
R.~D. {Blandford}, C.~F. {McKee}.
\newblock {Fluid dynamics of relativistic blast waves}.
\newblock Physics of Fluids 19 (1976) 1130--1138.

\bibitem{rees98}
M.~J. {Rees}, P.~{M\'esz\'aros}.
\newblock {Refreshed Shocks and Afterglow Longevity in Gamma-Ray Bursts}.
\newblock \apjl 496 (1998) L1--L4.

\bibitem{nishikawa05}
K.-I. {Nishikawa}, et~al.
\newblock {Particle Acceleration and Magnetic Field Generation in
  Electron-Positron Relativistic Shocks}.
\newblock \apj 622 (2005) 927--937.

\bibitem{spitkovsky08}
A.~{Spitkovsky}.
\newblock {Particle Acceleration in Relativistic Collisionless Shocks: Fermi
  Process at Last?}
\newblock \apjl 682 (2008) L5--L8.

\bibitem{nishikawa09}
K.-I. {Nishikawa}, et~al.
\newblock {Weibel Instability and Associated Strong Fields in a Fully
  Three-Dimensional Simulation of a Relativistic Shock}.
\newblock \apjl 698 (2009) L10--L13.

\bibitem{kennel84}
C.~F. {Kennel}, F.~V. {Coroniti}.
\newblock {Confinement of the Crab pulsar's wind by its supernova remnant}.
\newblock \apj 283 (1984) 694--709.

\bibitem{sironi09a}
L.~{Sironi}, A.~{Spitkovsky}.
\newblock {Particle Acceleration in Relativistic Magnetized Collisionless Pair
  Shocks: Dependence of Shock Acceleration on Magnetic Obliquity}.
\newblock \apj 698 (2009) 1523--1549.

\bibitem{sweet58}
P.~A. {Sweet}.
\newblock {The Neutral Point Theory of Solar Flares}.
\newblock In {B.~Lehnert}, editor, Electromagnetic Phenomena in Cosmical
  Physics, volume~6 of IAU Symposium (1958)\hspace{0pt} 123--134.

\bibitem{parker57}
E.~N. {Parker}.
\newblock {Sweet's Mechanism for Merging Magnetic Fields in Conducting Fluids}.
\newblock \jgr 62 (1957) 509--520.

\bibitem{thompson94}
C.~{Thompson}.
\newblock {A Model of Gamma-Ray Bursts}.
\newblock \mnras 270 (1994) 480--+.

\bibitem{rees05}
M.~J. {Rees}, P.~{M{\'e}sz{\'a}ros}.
\newblock {Dissipative Photosphere Models of Gamma-Ray Bursts and X-Ray
  Flashes}.
\newblock \apj 628 (2005) 847--852.

\bibitem{thompson07}
C.~{Thompson}, P.~{M{\'e}sz{\'a}ros}, M.~J. {Rees}.
\newblock {Thermalization in Relativistic Outflows and the Correlation between
  Spectral Hardness and Apparent Luminosity in Gamma-Ray Bursts}.
\newblock \apj 666 (2007) 1012--1023.

\bibitem{lazarian99}
A.~{Lazarian}, E.~T. {Vishniac}.
\newblock {Reconnection in a Weakly Stochastic Field}.
\newblock \apj 517 (1999) 700--718.

\bibitem{liangedison09}
E.~{Liang}, K.~{Noguchi}.
\newblock {Radiation from Comoving Poynting Flux Acceleration}.
\newblock \apj 705 (2009) 1473--1480.

\bibitem{smolsky96}
M.~V. {Smolsky}, V.~V. {Usov}.
\newblock {Relativistic Beam--Magnetic Barrier Collision and Nonthermal
  Radiation of Cosmological {$\gamma$}-Ray Bursters}.
\newblock \apj 461 (1996) 858--871.

\bibitem{ng06}
J.~S.~T. {Ng}, R.~J. {Noble}.
\newblock {Inductive and Electrostatic Acceleration in Relativistic Jet-Plasma
  Interactions}.
\newblock Physical Review Letters 96 (2006) 11 115006--+.

\bibitem{lemoine10}
M.~{Lemoine}, G.~{Pelletier}.
\newblock {Particle acceleration in GRBs}.
\newblock This volume.

\bibitem{band93}
D.~{Band}, et~al.
\newblock {BATSE observations of gamma-ray burst spectra. I - Spectral
  diversity}.
\newblock \apj 413 (1993) 281--292.

\bibitem{meszaros94}
P.~{M\'esz\'aros}, M.~J. {Rees}, H.~{Papathanassiou}.
\newblock {Spectral properties of blast-wave models of gamma-ray burst
  sources}.
\newblock \apj 432 (1994) 181--193.

\bibitem{tavani96}
M.~{Tavani}.
\newblock {A Shock Emission Model for Gamma-Ray Bursts. II. Spectral
  Properties}.
\newblock \apj 466 (1996) 768--778.

\bibitem{medvedev99}
M.~V. {Medvedev}, A.~{Loeb}.
\newblock {Generation of Magnetic Fields in the Relativistic Shock of Gamma-Ray
  Burst Sources}.
\newblock \apj 526 (1999) 697--706.

\bibitem{ghisellini00}
G.~{Ghisellini}, A.~{Celotti}, D.~{Lazzati}.
\newblock {Constraints on the emission mechanisms of gamma-ray bursts}.
\newblock \mnras 313 (2000) L1--L5.

\bibitem{peerzhang06}
A.~{Pe'er}, B.~{Zhang}.
\newblock {Synchrotron Emission in Small-Scale Magnetic Fields as a Possible
  Explanation for Prompt Emission Spectra of Gamma-Ray Bursts}.
\newblock \apj 653 (2006) 454--461.

\bibitem{asano09}
K.~{Asano}, T.~{Terasawa}.
\newblock {Slow Heating Model of Gamma-ray Burst: Photon Spectrum and 
Delayed Emission}.
\newblock \apj 705 (2009) 1714--1720.

\bibitem{daigne10}
F.~{Daigne}, Z.~{Bosnjak}, G.~{Dubus}.
\newblock {Reconciling observed GRB prompt spectra with synchrotron radiation?}
\newblock \aap 526 (2011) 110.


\bibitem{preece98}
R.~D. {Preece}, et~al.
\newblock {The Synchrotron Shock Model Confronts a ``Line of Death'' in the
  BATSE Gamma-Ray Burst Data}.
\newblock \apjl 506 (1998) L23--L26.

\bibitem{meszarosrees00}
P.~{M{\'e}sz{\'a}ros}, M.~J. {Rees}.
\newblock {Steep Slopes and Preferred Breaks in Gamma-Ray Burst Spectra: The
  Role of Photospheres and Comptonization}.
\newblock \apj 530 (2000) 292--298.

\bibitem{lloyd00}
N.~M. {Lloyd}, V.~{Petrosian}.
\newblock {Synchrotron Radiation as the Source of Gamma-Ray Burst Spectra}.
\newblock \apj 543 (2000) 722--732.

\bibitem{medvedev00}
M.~V. {Medvedev}.
\newblock {Theory of ``Jitter'' Radiation from Small-Scale Random Magnetic
  Fields and Prompt Emission from Gamma-Ray Burst Shocks}.
\newblock \apj 540 (2000) 704--714.

\bibitem{daigne98}
F.~{Daigne}, R.~{Mochkovitch}.
\newblock {Gamma-ray bursts from internal shocks in a relativistic wind:
  temporal and spectral properties}.
\newblock \mnras 296 (1998) 275--286.

\bibitem{bykov96}
A.~M. {Bykov}, P.~{Meszaros}.
\newblock {Electron Acceleration and Efficiency in Nonthermal Gamma-Ray
  Sources}.
\newblock \apjl 461 (1996) L37--L40.

\bibitem{medvedev06}
M.~V. {Medvedev}.
\newblock {Electron Acceleration in Relativistic Gamma-Ray Burst Shocks}.
\newblock \apjl 651 (2006) L9--L11.

\bibitem{sironi09b}
L.~{Sironi}, A.~{Spitkovsky}.
\newblock {Synthetic Spectra from Particle-In-Cell Simulations of Relativistic
  Collisionless Shocks}.
\newblock \apjl 707 (2009) L92--L96.

\bibitem{zenitani08}
S.~{Zenitani}, M.~{Hesse}.
\newblock {The role of the Weibel instability at the reconnection jet
front in relativistic pair plasma reconnection}.
\newblock Phys. Plasma 15 (2008) 022101.

\bibitem{panaitescu00b}
A.~{Panaitescu}, P.~{M{\'e}sz{\'a}ros}.
\newblock {Gamma-Ray Bursts from Upscattered Self-absorbed Synchrotron
  Emission}.
\newblock \apjl 544 (2000) L17--L21.

\bibitem{bosnjak09}
Z.~{Bosnjak}, D.~{Daigne}, G.~{Dubus}.
\newblock {Prompt high-energy emission from gamma-ray bursts in the 
 internal shock model}.
\newblock \aap 498 (2009) 677--703.

\bibitem{kumarnarayan09}
P.~{Kumar}, R.~{Narayan}.
\newblock {GRB 080319B: evidence for relativistic turbulence, not internal
  shocks}.
\newblock \mnras 395 (2009) 472--489.

\bibitem{derishev01}
E.~V. {Derishev}, V.~V. {Kocharovsky}, V.~V. {Kocharovsky}.
\newblock {Physical parameters and emission mechanism in gamma-ray bursts}.
\newblock \aap 372 (2001) 1071--1077.

\bibitem{piran09}
T.~{Piran}, R.~{Sari}, Y.-C. {Zou}.
\newblock {Observational limits on inverse Compton processes in gamma-ray
  bursts}.
\newblock \mnras 393 (2009) 1107--1113.

\bibitem{beskin10}
G.~{Beskin} et~al.
\newblock {Fast Optical Variability of a Naked-eye Burst -- Manifestation of 
the Periodic Activity of an Internal Engine}.
\newblock \apj 719 (2010) L10--L14.

\bibitem{peer06}
A.~{Pe'er}, P.~{M\'esz\'aros}, M. J.~{Rees}.
\newblock {The Observable Effects of a Photospheric Component on GRB and XRF 
Prompt Emission Spectrum}.
\newblock \apj 642 (2006) 995--1003.


\bibitem{thompson06}
C.~{Thompson}.
\newblock {Deceleration of a Relativistic, Photon-rich Shell: End of
  Preacceleration, Damping of Magnetohydrodynamic Turbulence, and the Emission
  Mechanism of Gamma-Ray Bursts}.
\newblock \apj 651 (2006) 333--365.

\bibitem{giannios08}
D.~{Giannios}.
\newblock {Prompt GRB emission from gradual energy dissipation}.
\newblock \aap 480 (2008) 305--312.

\bibitem{peerryde10}
A.~{Pe'er}, F.~{Ryde}.
\newblock {A Theory of Mulicolor Black Body Emission from Relativistically 
Expanding Plasmas}.
\newblock \apj submitted (2010) arXiv:1008.4590.


\bibitem{lazzati00}
D.~{Lazzati}, et~al.
\newblock {Compton-dragged Gamma-Ray Bursts Associated with Supernovae}.
\newblock \apjl 529 (2000) L17--L20.

\bibitem{dar04}
A.~{Dar}, A.~{de R{\'u}jula}.
\newblock {Towards a complete theory of gamma-ray bursts}.
\newblock \physrep 405 (2004) 203--278.

\bibitem{wang07}
X.-Y. {Wang}, et~al.
\newblock {Nonthermal Gamma-Ray/X-Ray Flashes from Shock Breakout in Gamma-Ray
  Burst-Associated Supernovae}.
\newblock \apj 664 (2007) 1026--1032.

\bibitem{gupta07}
N.~{Gupta}, B.~{Zhang}.
\newblock {Prompt emission of high-energy photons from gamma ray bursts}.
\newblock \mnras 380 (2007) 78--92.

\bibitem{asano09b}
K.~{Asano}, S.~{Inoue}, P.~{M\'esz\'aros}.
\newblock {Prompt high-energy emission from proton-dominated gamma-ray bursts}.
\newblock \apj 699 (2009) 953--957.

\bibitem{razzaque10}
S.~{Razzaque}, C.~D. {Dermer}, J.~D. {Fink}.
\newblock {Synchrotron Radiation from Ultra-High Energy Protons and the Fermi 
Observations of GRB 080916C}.
\newblock Open Astron. J. 1 (2010) 150--155.

\bibitem{shaviv95}
N.~J. {Shaviv}, A.~{Dar}.
\newblock {Gamma-Ray Bursts from Minijets}.
\newblock \apj 447 (1995) 863--873.

\bibitem{lazzati04}
D.~{Lazzati}, et~al.
\newblock {Compton drag as a mechanism for very high linear polarization in
  gamma-ray bursts}.
\newblock \mnras 347 (2004) L1--L5.


\bibitem{atteia10}
{J. L.}~{Atteia}, M.~{Boer}.
\newblock {Catching the prompt emission from GRBs}.
\newblock This volume.


\bibitem{burrows05}
D.~N. {Burrows}, et~al.
\newblock {Bright X-ray Flares in Gamma-Ray Burst Afterglows}.
\newblock Science 309 (2005) 1833--1835.

\bibitem{chincarini07}
G.~{Chincarini}, et~al.
\newblock {The First Survey of X-Ray Flares from Gamma-Ray Bursts Observed by
  Swift: Temporal Properties and Morphology}.
\newblock \apj 671 (2007) 1903--1920.

\bibitem{falcone07}
A.~D. {Falcone}, et~al.
\newblock {The First Survey of X-Ray Flares from Gamma-Ray Bursts Observed by
  Swift: Spectral Properties and Energetics}.
\newblock \apj 671 (2007) 1921--1938.

\bibitem{ioka05}
K.~{Ioka}, S.~{Kobayashi}, B.~{Zhang}.
\newblock {Variabilities of Gamma-Ray Burst Afterglows: Long-acting Engine,
  Anisotropic Jet, or Many Fluctuating Regions?}
\newblock \apj 631 (2005) 429--434.

\bibitem{fanwei05}
Y.~Z. {Fan}, D.~M. {Wei}.
\newblock {Late internal-shock model for bright X-ray flares in gamma-ray burst
  afterglows and GRB 011121}.
\newblock \mnras 364 (2005) L42--L46.

\bibitem{lazzati07}
D.~{Lazzati}, R.~{Perna}.
\newblock {X-ray flares and the duration of engine activity in gamma-ray
  bursts}.
\newblock \mnras 375 (2007) L46--L50.

\bibitem{maxham09}
A.~{Maxham}, B.~{Zhang}.
\newblock {Modeling Gamma-Ray Burst X-Ray Flares Within the Internal Shock
  Model}.
\newblock \apj 707 (2009) 1623--1633.

\bibitem{beloborodov10b}
A.~M. {Beloborodov} et~al.
\newblock {Is GRB afterglow emission intrinsically anisotropic?}
\newblock \mnras 410 (2011) 2422--2427.


\bibitem{sarimeszaros00}
R.~{Sari}, P.~{M{\'e}sz{\'a}ros}.
\newblock {Impulsive and Varying Injection in Gamma-Ray Burst Afterglows}.
\newblock \apjl 535 (2000) L33--L37.

\bibitem{ghisellini07}
G.~{Ghisellini}, et~al.
\newblock {``Late Prompt'' Emission in Gamma-Ray Bursts?}
\newblock \apjl 658 (2007) L75--L78.

\bibitem{kumar08b}
P.~{Kumar}, R.~{Narayan}, J.~L. {Johnson}.
\newblock {Mass fall-back and accretion in the central engine of gamma-ray
  bursts}.
\newblock \mnras 388 (2008) 1729--1742.

\bibitem{cannizzo09}
J.~K. {Cannizzo}, N.~{Gehrels}.
\newblock {A New Paradigm for Gamma-ray Bursts: Long-term Accretion Rate
  Modulation by an External Accretion Disk}.
\newblock \apj 700 (2009) 1047--1058.

\bibitem{lindner10}
C.~C. {Lindner}, et~al.
\newblock {Collapsar Accretion and the Gamma-Ray Burst X-Ray Light Curve}.
\newblock \apj 713 (2010) 800--815.

\bibitem{panai06b}
A.~{Panaitescu}, et~al.
\newblock {Evidence for chromatic X-ray light-curve breaks in Swift gamma-ray
  burst afterglows and their theoretical implications}.
\newblock \mnras 369 (2006) 2059--2064.

\bibitem{liang06b}
E.-W. {Liang}, et~al.
\newblock {Temporal Profiles and Spectral Lags of XRF 060218}.
\newblock \apjl 653 (2006) L81--L84.

\bibitem{depasquale09}
M.~{de Pasquale}, et~al.
\newblock {Jet breaks at the end of the slow decline phase of Swift GRB light
  curves}.
\newblock \mnras 392 (2009) 153--169.

\bibitem{genet07}
F.~{Genet}, F.~{Daigne}, R.~{Mochkovitch}.
\newblock {Can the early X-ray afterglow of gamma-ray bursts be explained by a
  contribution from the reverse shock?}
\newblock \mnras 381 (2007) 732--740.

\bibitem{uhm07}
Z.~L. {Uhm}, A.~M. {Beloborodov}.
\newblock {On the Mechanism of Gamma-Ray Burst Afterglows}.
\newblock \apjl 665 (2007) L93--L96.

\bibitem{shao05}
L.~{Shao}, Z.~G. {Dai}.
\newblock {A Reverse-Shock Model for the Early Afterglow of GRB 050525A}.
\newblock \apj 633 (2005) 1027--1030.

\bibitem{shen09b}
{R.-F.} {Shen}, et~al.
\newblock {The dust scattering model cannot explain the shallow X-ray decay in
  GRB afterglows}.
\newblock \mnras 393 (2009) 598--606.

\bibitem{wu10}
{X.-F.} {Wu}, B.~{Zhang}.
\newblock {X-ray afterglow from photosphere of a long lasting engine-driven wind}.
\newblock \apj submitted (2011).


\bibitem{king05}
A.~{King}, et~al.
\newblock {Gamma-Ray Bursts: Restarting the Engine}.
\newblock \apjl 630 (2005) L113--L115.

\bibitem{perna06}
R.~{Perna}, P.~J. {Armitage}, B.~{Zhang}.
\newblock {Flares in Long and Short Gamma-Ray Bursts: A Common Origin in a
  Hyperaccreting Accretion Disk}.
\newblock \apjl 636 (2006) L29--L32.

\bibitem{proga06}
D.~{Proga}, B.~{Zhang}.
\newblock {The late time evolution of gamma-ray bursts: ending hyperaccretion
  and producing flares}.
\newblock \mnras 370 (2006) L61--L65.

\bibitem{dai06}
Z.~G. {Dai}, et~al.
\newblock {X-ray Flares from Postmerger Millisecond Pulsars}.
\newblock Science 311 (2006) 1127--1129.

\bibitem{lee09}
W.~H. {Lee}, E.~{Ramirez-Ruiz}, D.~{L{\'o}pez-C{\'a}mara}.
\newblock {Phase Transitions and He-Synthesis-Driven Winds in Neutrino Cooled
  Accretion Disks: Prospects for Late Flares in Short Gamma-Ray Bursts}.
\newblock \apjl 699 (2009) L93--L96.

\bibitem{popham99}
R.~{Popham}, S.~E. {Woosley}, C.~{Fryer}.
\newblock {Hyperaccreting Black Holes and Gamma-Ray Bursts}.
\newblock \apj 518 (1999) 356--374.

\bibitem{yuan10}
F.~{Yuan} et~al.
\newblock in preparation (2010).

\bibitem{fan05e}
Y.~Z. {Fan}, B.~{Zhang}, D.~{Proga}.
\newblock {Linearly Polarized X-Ray Flares following Short Gamma-Ray Bursts}.
\newblock \apjl 635 (2005) L129--L132.

\bibitem{macfadyen01}
A.~I. {MacFadyen}, S.~E. {Woosley}, A.~{Heger}.
\newblock {Supernovae, Jets, and Collapsars}.
\newblock \apj 550 (2001) 410--425.

\bibitem{rosswog07}
S.~{Rosswog}.
\newblock {Fallback accretion in the aftermath of a compact binary merger}.
\newblock \mnras 376 (2007) L48--L51.

\bibitem{lazzati08}
D.~{Lazzati}, R.~{Perna}, M.~C. {Begelman}.
\newblock {X-ray flares, neutrino-cooled discs and the dynamics of late
  accretion in gamma-ray burst engines}.
\newblock \mnras 388 (2008) L15--L19.

\bibitem{corsi09}
A.~{Corsi}, P.~{M{\'e}sz{\'a}ros}.
\newblock {Gamma-ray Burst Afterglow Plateaus and Gravitational Waves:
  Multi-messenger Signature of a Millisecond Magnetar?}
\newblock \apj 702 (2009) 1171--1178.

\bibitem{sari98}
R.~{Sari}, T.~{Piran}, R.~{Narayan}.
\newblock {Spectra and Light Curves of Gamma-Ray Burst Afterglows}.
\newblock \apjl 497 (1998) L17+.

\bibitem{dailu98c}
Z.~G. {Dai}, T.~{Lu}.
\newblock {Gamma-ray burst afterglows: effects of radiative corrections and
  non-uniformity of the surrounding medium}.
\newblock \mnras 298 (1998) 87--92.

\bibitem{chevalier00}
R.~A. {Chevalier}, Z.-Y. {Li}.
\newblock {Wind Interaction Models for Gamma-Ray Burst Afterglows: The Case for
  Two Types of Progenitors}.
\newblock \apj 536 (2000) 195--212.

\bibitem{wijers99}
R.~A.~M.~J. {Wijers}, T.~J. {Galama}.
\newblock {Physical Parameters of GRB 970508 and GRB 971214 from Their
  Afterglow Synchrotron Emission}.
\newblock \apj 523 (1999) 177--186.

\bibitem{panaitescu01}
A.~{Panaitescu}, P.~{Kumar}.
\newblock {Fundamental Physical Parameters of Collimated Gamma-Ray Burst
  Afterglows}.
\newblock \apjl 560 (2001) L49--L53.

\bibitem{panaitescu02}
A.~{Panaitescu}, P.~{Kumar}.
\newblock {Properties of Relativistic Jets in Gamma-Ray Burst Afterglows}.
\newblock \apj 571 (2002) 779--789.

\bibitem{yost03}
S.~A. {Yost}, et~al.
\newblock {A Study of the Afterglows of Four Gamma-Ray Bursts: Constraining the
  Explosion and Fireball Model}.
\newblock \apj 597 (2003) 459--473.

\bibitem{rhoads99}
J.~E. {Rhoads}.
\newblock {The Dynamics and Light Curves of Beamed Gamma-Ray Burst Afterglows}.
\newblock \apj 525 (1999) 737--749.

\bibitem{sari99}
R.~{Sari}, T.~{Piran}, J.~P. {Halpern}.
\newblock {Jets in Gamma-Ray Bursts}.
\newblock \apjl 519 (1999) L17--L20.

\bibitem{zhangmeszaros02a}
B.~{Zhang}, P.~{M{\'e}sz{\'a}ros}.
\newblock {Gamma-Ray Bursts with Continuous Energy Injection and Their
  Afterglow Signature}.
\newblock \apj 566 (2002) 712--722.

\bibitem{meszaros98}
P.~{M\'esz\'aros}, M.~J. {Rees}, R.~A.~M.~J. {Wijers}.
\newblock {Viewing Angle and Environment Effects in Gamma-Ray Bursts: Sources
  of Afterglow Diversity}.
\newblock \apj 499 (1998) 301--308.

\bibitem{zhangmeszaros02b}
B.~{Zhang}, P.~{M{\'e}sz{\'a}ros}.
\newblock {Gamma-Ray Burst Beaming: A Universal Configuration with a Standard
  Energy Reservoir?}
\newblock \apj 571 (2002) 876--879.

\bibitem{rossi02}
E.~{Rossi}, D.~{Lazzati}, M.~J. {Rees}.
\newblock {Afterglow light curves, viewing angle and the jet structure of
  {$\gamma$}-ray bursts}.
\newblock \mnras 332 (2002) 945--950.

\bibitem{huang99}
Y.~F. {Huang}, Z.~G. {Dai}, T.~{Lu}.
\newblock {A generic dynamical model of gamma-ray burst remnants}.
\newblock \mnras 309 (1999) 513--516.

\bibitem{huang03}
Y.~F. {Huang}, K.~S. {Cheng}.
\newblock {Gamma-ray bursts: optical afterglows in the deep Newtonian phase}.
\newblock \mnras 341 (2003) 263--269.

\bibitem{panai06a}
A.~{Panaitescu}, et~al.
\newblock {Analysis of the X-ray emission of nine Swift afterglows}.
\newblock \mnras 366 (2006) 1357--1366.

\bibitem{grupe07}
D.~{Grupe}, et~al.
\newblock {Swift and XMM-Newton Observations of the Extraordinary Gamma-Ray
  Burst 060729: More than 125 Days of X-Ray Afterglow}.
\newblock \apj 662 (2007) 443--458.

\bibitem{mangano07b}
V.~{Mangano}, et~al.
\newblock {Swift observations of GRB 060614: an anomalous burst with a well
  behaved afterglow}.
\newblock \aap 470 (2007) 105--118.

\bibitem{berger03}
E.~{Berger}, S.~R. {Kulkarni}, D.~A. {Frail}.
\newblock {A Standard Kinetic Energy Reservoir in Gamma-Ray Burst Afterglows}.
\newblock \apj 590 (2003) 379--385.

\bibitem{dai08}
X.~{Dai} et~al.
\newblock {Go Long, Go Deep: Finding Optical Jet Breaks for Swift-Era GRBs with the LBT}.
\newblock \apj 682 (2008) L77--L80.


\bibitem{kobayashizhang03a}
S.~{Kobayashi}, B.~{Zhang}.
\newblock {GRB 021004: Reverse Shock Emission}.
\newblock \apjl 582 (2003) L75--L78.

\bibitem{akerlof99}
C.~{Akerlof}, et~al.
\newblock {Observation of contemporaneous optical radiation from a
  {$\gamma$}-ray burst}.
\newblock \nat 398 (1999) 400--402.

\bibitem{jinfan07}
Z.~P. {Jin}, Y.~Z. {Fan}.
\newblock {GRB 060418 and 060607A: the medium surrounding the progenitor and
  the weak reverse shock emission}.
\newblock \mnras 378 (2007) 1043--1048.

\bibitem{wu03}
X.~F. {Wu}, et~al.
\newblock {Optical flashes and very early afterglows in wind environments}.
\newblock \mnras 342 (2003) 1131--1138.

\bibitem{kobayashizhang03b}
S.~{Kobayashi}, B.~{Zhang}.
\newblock {Early Optical Afterglows from Wind-Type Gamma-Ray Bursts}.
\newblock \apj 597 (2003) 455--458.

\bibitem{kobayashi07}
S.~{Kobayashi}, et~al.
\newblock {Inverse Compton X-Ray Flare from Gamma-Ray Burst Reverse Shock}.
\newblock \apj 655 (2007) 391--395.

\bibitem{wang01}
X.~Y. {Wang}, Z.~G. {Dai}, T.~{Lu}.
\newblock {Prompt High-Energy Gamma-Ray Emission from the Synchrotron
  Self-Compton Process in the Reverse Shocks of Gamma-Ray Bursts}.
\newblock \apjl 546 (2001) L33--L37.

\bibitem{wang01b}
X.~Y. {Wang}, Z.~G. {Dai}, T.~{Lu}.
\newblock {The Inverse Compton Emission Spectra in the Very Early Afterglows of
  Gamma-Ray Bursts}.
\newblock \apj 556 (2001) 1010--1016.

\bibitem{godet10}
O.~{Godet}, R.~{Mochkovitch}.
\newblock {Afterglows after Swift}.
\newblock This volume.


\bibitem{hurley94}
K.~{Hurley}, et~al.
\newblock {Detection of a Gamma-Ray Burst of Very Long Duration and Very High
  Energy}.
\newblock \nat 372 (1994) 652--654.

\bibitem{gonzalez03}
M.~M. {Gonz{\'a}lez}, et~al.
\newblock {A {$\gamma$}-ray burst with a high-energy spectral component
  inconsistent with the synchrotron shock model}.
\newblock \nat 424 (2003) 749--751.

\bibitem{piron10}
F.~{Piron}, V.~{Connaughton}.
\newblock {The Fermi view of GRBs}.
\newblock This volume.


\bibitem{ackermann11}
M. {Ackermann}, et~al.
\newblock {Detection of a Spectral Break in the Extra Hard Component of GRB 090926A}.
\newblock \apj 729 (2011) 114.


\bibitem{ghisellini10}
G.~{Ghisellini}, et~al.
\newblock {GeV emission from gamma-ray bursts: a radiative fireball?}
\newblock \mnras 403 (2010) 926--937.

\bibitem{maxham11}
A.~{Maxham}, {B.-B.} {Zhang}, B.~{Zhang}.
\newblock {Is GeV Emission from Gamma-Ray Bursts of External Shock Origin?}.
\newblock \mnras in press (2011), arXiv:1101.0144.


\bibitem{toma10}
K.~{Toma}, {X.-F.} {Wu}, P.~{M{\'e}sz{\'a}ros}.
\newblock {A Photosphere-Internal Shock Model of Gamma-Ray Bursts: Implications 
for the Fermi/LAT Results}.
\newblock \mnras submitted (2010), arXiv:1002.2634.


\bibitem{kumar09}
P.~{Kumar}, R.~{Barniol Duran}.
\newblock {On the generation of high-energy photons detected by the Fermi
  Satellite from gamma-ray bursts}.
\newblock \mnras 400 (2009) L75--L79.

\bibitem{kumar10}
P.~{Kumar}, R.~{Barniol Duran}.
\newblock {External forward shock origin of high-energy emission for three 
gamma-ray bursts detected by Fermi}.
\newblock \mnras 409 (2010) 226--236.

\bibitem{toma09}
K.~{Toma}, {X.-F.} {Wu}, P.~{M{\'e}sz{\'a}ros}.
\newblock {An Up-Scattered Cocoon Emission Model of Gamma-Ray Burst High-Energy
  Lags}.
\newblock \apj 707 (2009) 1404--1416.

\bibitem{wang10}
{X.-Y.}~{Wang} et~al.
\newblock {Klein-Nishina Effects on the High-energy Afterglow Emission of Gamma-ray Bursts}.
\newblock \apj 712 (2010) 1232--1240.

\bibitem{feng10}
S. Y. {Feng}, Z. G. {Dai}
\newblock {Multiband Fitting to Three Long GRBs with Fermi/LAT Data: 
Structured Ejecta Sweeping up a Density-Jump Medium}
\newblock \apj submitted (2010) arXiv:1011.3103.


\bibitem{he10}
{H.-N.}~{He} et~al.
\newblock {On the High Energy Emission of the Short GRB 090510}.
\newblock \apj in press (2011), arXiv:1009.1432.

\bibitem{liu10}
{R.-Y.}~{Liu}, {X.-Y.}~Wang.
\newblock {Modeling the broadband emission of Fermi/LAT GRB 090902B}.
\newblock \apj 730 (2011) 1.

\bibitem{li10}
Z.~{Li}.
\newblock {Prompt GeV Emission from Residual Collisions in Gamma-Ray Burst 
Outflows: Evidence from Fermi Observations of GRB 080916C}.
\newblock \apj 709 (2010) 525--534.

\bibitem{ioka10}
K.~{Ioka}.
\newblock {Very High Lorentz Factor Fireballs and Gamma-Ray Burst Spectra}.
\newblock Prog. Theor. Phys. 124 (2010) 667--710.

\bibitem{depasquale10}
M.~{De Pasquale}, et~al.
\newblock {Swift and Fermi Observations of the Early Afterglow of the Short
  Gamma-Ray Burst 090510}.
\newblock \apjl 709 (2010) L146--L151.

\bibitem{hopkins06}
A.~M. {Hopkins}, J.~F. {Beacom}.
\newblock {On the Normalization of the Cosmic Star Formation History}.
\newblock \apj 651 (2006) 142--154.

\bibitem{kistler08}
M.~D. {Kistler}, et~al.
\newblock {An Unexpectedly Swift Rise in the Gamma-Ray Burst Rate}.
\newblock \apjl 673 (2008) L119--L122.

\bibitem{bromm06}
V.~{Bromm}, A.~{Loeb}.
\newblock {High-Redshift Gamma-Ray Bursts from Population III Progenitors}.
\newblock \apj 642 (2006) 382--388.

\bibitem{butler10}
N.~R. {Butler}, J.~S. {Bloom}, D.~{Poznanski}.
\newblock {The Cosmic Rate, Luminosity Function, and Intrinsic Correlations of
  Long Gamma-Ray Bursts}.
\newblock \apj 711 (2010) 495--516.

\bibitem{qin10}
{S.-F.} {Qin}, et~al.
\newblock {Simulations on high-z long gamma-ray burst rate}.
\newblock \mnras 406 (2010) 558--565.

\bibitem{virgili10b}
F.~{Virgili}, et~al.
\newblock in preparation (2010).

\bibitem{campisi09}
M.~A. {Campisi}, et~al.
\newblock {Properties of long gamma-ray burst host galaxies in cosmological
  simulations}.
\newblock \mnras 400 (2009) 1613--1624.

\bibitem{holder03}
G.~P. {Holder}, et~al.
\newblock {The Reionization History at High Redshifts. II. Estimating the
  Optical Depth to Thomson Scattering from Cosmic Microwave Background
  Polarization}.
\newblock \apj 595 (2003) 13--18.

\bibitem{fanx06}
X.~{Fan}, C.~L. {Carilli}, B.~{Keating}.
\newblock {Observational Constraints on Cosmic Reionization}.
\newblock \araa 44 (2006) 415--462.

\bibitem{totani06}
T.~{Totani}, et~al.
\newblock {Implications for Cosmic Reionization from the Optical Afterglow
  Spectrum of the Gamma-Ray Burst 050904 at z = 6.3}.
\newblock \pasj 58 (2006) 485--498.

\bibitem{nagamine08}
K.~{Nagamine}, B.~{Zhang}, L.~{Hernquist}.
\newblock {Incidence Rate of GRB-Host DLAs at High Redshift}.
\newblock \apjl 686 (2008) L57--L60.

\bibitem{pontzen10}
A.~{Pontzen} et al.
\newblock {The nature of HI absorbers in gamma-ray burst afterglows: 
clues from hydrodynamic simulations}.
\newblock \mnras 402 (2010) 1523--1535.


\bibitem{abel02}
T.~{Abel}, G.~L. {Bryan}, M.~L. {Norman}.
\newblock {The Formation of the First Star in the Universe}.
\newblock Science 295 (2002) 93--98.

\bibitem{komissarov10}
S.~S. {Komissarov}, M.~V. {Barkov}.
\newblock {Supercollapsars and their X-ray bursts}.
\newblock \mnras 402 (2010) L25--L29.

\bibitem{meszarosrees10}
P.~{M{\'e}sz{\'a}ros}, M.~J. {Rees}.
\newblock {Population III Gamma-ray Bursts}.
\newblock \apj 715 (2010) 967--971.

\bibitem{turk09}
M.~J. {Turk}, T.~{Abel}, B.~{O'Shea}.
\newblock {The Formation of Population III Binaries from Cosmological Initial
  Conditions}.
\newblock Science 325 (2009) 601--603.

\bibitem{janiuk08a}
A.~{Janiuk}, D.~{Proga}.
\newblock {Low Angular Momentum Accretion in the Collapsar: How Long Can a Long
  GRB Be?}
\newblock \apj 675 (2008) 519--527.

\bibitem{janiuk08b}
A.~{Janiuk}, R.~{Moderski}, D.~{Proga}.
\newblock {On the Duration of Long GRBs: Effects of Black Hole Spin}.
\newblock \apj 687 (2008) 433--442.

\bibitem{petitjean10}
P.~{Petitjean}.
\newblock {GRBs as probes of the distant universe}.
\newblock This volume.


\bibitem{yamazaki09}
R.~{Yamazaki}.
\newblock {Prior Emission Model for X-ray Plateau Phase of Gamma-Ray Burst Afterglows}.
\newblock \apj 690 (2009) L118-L121.


\bibitem{liang09}
{E.-W.} Liang et~al.
\newblock {A Comprehensive Analysis of Swift/X-Ray Telescope Data. IV. Single Power-Law 
Decaying Light Curves Versus Canonical Light Curves and Implications for a Unified 
Origin of X-Rays}.
\newblock \apj 707 (2009) 328--342.


\end{thebibliography}


\end{document}